\documentclass[amsmath,amssymb,twocolumn,superscriptaddress,preprintnumbers,nofootinbib,prd]{revtex4-1}

\usepackage[utf8]{inputenc}
\usepackage{amssymb}
\usepackage{cancel}
\usepackage{amsmath}
\usepackage{braket}

\input{header}

\begin{document}


\vspace*{-0.7cm}

\title{Dark Vector Boson Bremsstrahlung: New Form Factors for a Broader Class of Models}

\author{F. Kling}
\email{fkling@uci.edu}
\affiliation{Department of Physics and Astronomy, University of California, Irvine, CA 92697, USA}
\affiliation{Deutsches Elektronen-Synchrotron DESY, Notkestr. 85, 22607 Hamburg, Germany}

\author{P. Reimitz}
 \email{peter@if.usp.br}
\affiliation{Instituto de F\'{i}sica,
Universidade de Sāo Paulo, 05508-090 Sāo Paulo, SP, Brasil}
\affiliation{Department of Physics and Astronomy, University of Victoria, Victoria BC V8P 5C2, Canada}

\author{A. Ritz}
 \email{aritz@uvic.ca}
\affiliation{Department of Physics and Astronomy, University of Victoria, Victoria BC V8P 5C2, Canada}


\begin{abstract}
We explore the sensitivity of collider experiments to a broad class of GeV-scale dark vector models of new physics via production in proton and neutron bremsstrahlung and initial state radiation. This is achieved using a new physically motivated model for timelike vector form factors with generic charges for both protons and neutrons, which is fit to a variety of timelike and spacelike data with quantified uncertainties. The production model for both proton and neutron bremsstrahlung is applied to re-cast and extend the reach of existing FASER data to GeV-mass dark photons, $U(1)_B$, $U(1)_{B-L}$, and photophobic vectors, as well as forecasts for millicharged particles at FORMOSA.  
\end{abstract}

\maketitle

\section{Introduction}
\label{sec:intro}

The nature of dark matter, the origin of the matter-antimatter asymmetry, and the existence of nonzero neutrino masses remain open questions in fundamental physics. Each points toward physics beyond the Standard Model (BSM) but is largely agnostic about the mass scale of the new degrees of freedom involved. Experimental efforts have progressively extended sensitivity to a wide range of possible masses for BSM particles, while theoretical work has both motivated and refined searches in specific regions of parameter space~\cite{Feng:2010gw,Alexander:2016aln,Bertone:2018krk,Arcadi:2017kky,Roszkowski:2017nbc}. 

The exploration of light states below the electroweak scale, interacting only through extremely weak couplings, is particularly compelling~\cite{Pospelov:2008zw,Batell:2009yf,Essig:2009nc,Reece:2009un,Freytsis:2009bh,Batell:2009jf,Freytsis:2009ct,Essig:2010xa,Essig:2010gu,McDonald:2010fe,Williams:2011qb,Davoudiasl:2012ag,Kahn:2012br,Andreas:2012mt,Battaglieri:2017aum,Knapen:2017xzo}. This dark sector framework focuses attention on a small number of renormalizable interactions - the scalar, vector, and neutrino portals - that could connect a neutral hidden sector to the Standard Model (SM). Such scenarios can be probed with near and far detectors at a diverse array of accelerator facilities~\cite{Batell:2009di,deNiverville:2011it,deNiverville:2012ij,Kahn:2014sra,Adams:2013qkq,Soper:2014ska,Dobrescu:2014ita,Coloma:2015pih,Alpigiani:2018fgd,Ariga:2018pin,Dutta:2020vop,Batell:2021blf,Batell:2021aja,Bjorken:2009mm,Izaguirre:2013uxa,Diamond:2013oda,Izaguirre:2014dua,Batell:2014mga,Lees:2017lec,Berlin:2020uwy,Krnjaic:2022ozp,Berlin:2018bsc,Bauer:2018onh,Berlin:2023qco,Lu:2023cet,Mongillo:2023hbs,Filimonova:2022pkj,CarrilloGonzalez:2021lxm,Batell:2021snh,Garcia:2024uwf} spanning a broad range of energies and experimental techniques~\cite{APEX:2011dww,KLOE-2:2011hhj,BaBar:2012bkw,Beacham:2019nyx,LBNE:2013dhi,MiniBooNE:2017nqe,MiniBooNEDM:2018cxm,MATHUSLA:2018bqv,BaBar:2017tiz,NA64:2019imj}. Searches in this regime are typically framed in terms of Feebly Interacting Particle (FIP) searches~\cite{Antel:2023hkf}.

A representative FIP benchmark is a new GeV-scale Abelian gauge boson $A'$, often called a dark photon, arises from a hidden $U(1)$ symmetry that mixes kinetically with hypercharge. Variants of this framework include a $U(1)_{B-L}$ gauge boson, which couples to baryon and lepton number, protophobic models, where couplings to protons are suppressed, and scenarios predicting millicharged particles with extremely small electric charges. Together, these scenarios illustrate the rich landscape of weakly coupled sectors accessible to current and upcoming searches. 

While larger couplings, leading to prompt decays, have been tightly constrained by detectors near the interaction point of collision experiments, current efforts increasingly focus on long-lived particle searches~\cite{Berlin:2018jbm}. Examples include forward experiments at colliders such as FASER at the LHC~\cite{Feng:2017uoz}, upcoming fixed-target facilities like SHiP at the CERN North Area~\cite{SHiP:2015vad}, as well as searches at high intensity neutrino experiments, such as those of the short baseline program at Fermilab~\cite{Machado:2019oxb,DUNE:2020fgq}.  

This motivates a careful examination of all production and detection modes~\cite{Celentano:2020vtu,Capozzi:2021nmp,Blinov:2024pza,LoChiatto:2024guj,Altmannshofer:2022ckw,Curtin:2023bcf,Kyselov:2024dmi}. In high-energy proton-proton collisions, Drell-Yan production is the least suppressed parton-level process and typically dominates the production of 10 GeV -- TeV dark photons. In the MeV regime, the copious production of pions and $\eta$-mesons - and their potential decays into new light states - make meson decays the leading source of dark photons. At intermediate masses around and above 1 GeV, dark bremsstrahlung emerges as the primary production mode. While this mass range remains challenging for existing experiments, it will be central for future facilities such as the Forward Physics Facility (FPF)~\cite{Anchordoqui:2021ghd, Feng:2022inv, Adhikary:2024nlv} and SHiP, making accurate modeling of this process particularly important. 

The approach of Ref.~\cite{Blumlein:2013cua} by Bl\"{u}mlein and Brunner has for some time provided the standard framework for modeling dark bremsstrahlung and therefore has been widely adopted in the literature~\cite{deNiverville:2016rqh,Feng:2017uoz}. Recently, however, this description has been revisited~\cite{Foroughi-Abari:2021zbm,Foroughi-Abari:2024xlj,Gorbunov:2024vrc,Gorbunov:2024iyu}, leading to the development of improved treatments. Current efforts focus on initial state radiation, refining splitting functions and incorporating more accurate form factor parameterizations, resulting in more precise predictions for $A'$ production rates and initiating discussions around a robust uncertainty estimate. 

In this work, we present a generic nucleon form-factor model based on a simple sum over vector meson resonances. This approach is designed to satisfy not only the SM normalization conditions at $t=0$, but also consistent normalization prescriptions for arbitrary light-quark couplings in BSM scenarios. To ensure flexibility for such BSM applications, we additionally impose superconvergence relations on each vector-meson contribution individually. By fitting both proton and neutron data simultaneously, we extend the formalism to neutron bremsstrahlung. 

The remainder of this paper is organized as follows. \cref{sec:models} reviews the models of new light vector bosons, while \cref{sec:ffs} derives the nucleon form-factor model. \cref{sec:modeling} describes our modeling of vector boson production via Bremsstrahlung. \cref{sec:FASER} utilizes this production model and presents improved predictions for dark vector detection, with particular focus on FASER sensitivities while \cref{sec:application} demonstrates how the framework can be applied to other BSM models. \cref{sec:conclusion} concludes with a summary and outlook.

\section{Models of Vector Bosons}
\label{sec:models}

\begin{table}[t]
\centering
\begin{tabular}{l|ccc|ccc|cc|cc}
\hline\hline
Model & $x_u$ & $x_d$ & $x_s$ & $x_\omega$ & $x_\rho$ & $x_\phi$& $x_p$ & $x_n$ & $x_\ell$ & $x_\nu$ \\
\hline\hline
Dark photon & $+2/3$ & $-1/3$ & $-1/3$ & 1 & 1 & 1 & 1&0&-1&0\\
$U(1)_{B-L}$ & $+1/3$ & $+1/3$ & $+1/3$ & 2 & 0  & -1 & 1&1&-1&-1\\
$U(1)_{B}$ & $+1/3$ & $+1/3$ & $+1/3$ & 2 & 0  & -1 & 1&1&0&0\\
Protophobic & $-1/3$ & $+2/3$ & $+2/3$  & 1  & -1  & -2& 0&1&-1&0\\
\hline\hline
\end{tabular}
\caption{Benchmark gauge boson models and the corresponding charges $x_f$ to the SM fermions, vector mesons and nucleons.}
\label{tab:charges}
\end{table}

In this study, we consider a generic model of a gauge boson $A'$ coupling to the SM fermions $f$. The interaction Lagrangian of this particle is given by
\be
\mathcal{L} \supset  g \sum_{f} x_f \bar{f} \gamma^\mu  f A'_{\mu} \ . 
\label{eq:Lint}
\ee
Here $g$ is the gauge coupling and $x_f$ are the charges of the fermions under the gauge group. 

The best known example of such a vector boson is a dark photon. It arises as a massive gauge boson of a broken $U(1)$ symmetry in a hidden sector, which kinetically mixes with the SM photon of the SM. As a result of this kinetic mixing, the dark photon couples to fermions proportional to their electric charge, so $x_f = q_f$. The gauge coupling \cref{eq:Lint} then takes the form $g = \epsilon e$, where $e$ the electron charge and $\epsilon$ is the kinetic mixing parameter. Since the dark photon has the same coupling structure as a SM photon, its coupling to nucleons is described by the same electromagnetic form factors $F_1(t)$ and $F_2(t)$, evaluated at $t =m_{A'}^2$ for on-shell dark photons. 

Alternatively, new gauge bosons are predicted if one of the global symmetries of the SM is gauged. Frequently studied examples include the gauge bosons of a U(1)$_{B-L}$ or U(1)$_B$ symmetry, where $B$ and $L$ correspond to the baryon number and total lepton number of the SM. Another example is a protophobic gauge boson, whose couplings to quarks are orthogonal to those of a dark photon. Under this group, effectively the role of the proton and neutron are exchanged in comparison to the dark photon case - a feature that we will use in the construction of a general form factor model. The charges $x_f$ of the SM fermions for the gauge boson models discussed in this study are  listed in \cref{tab:charges}.

Thus far, no form factors have been provided to describe the interaction of gauge bosons with different coupling structures to the nucleons. The following section will address this shortcoming and derive a general and flexible form factor model that can easily be applied to models of new vector bosons with arbitrary coupling structures. 

\section{General Nucleon Form Factors}
\label{sec:ffs}

\subsection{Normalization}

We follow the vector meson dominance framework and assume that all proton and neutron form factors $F_i^{p,n}=F_i^s\pm F_i^v$ with $i=1,2$ consisting of iso-scalar $F_i^s=F_{i,\omega} + F_{i,\phi}$ and iso-vector $F_i^v = F_{i,\rho}$ components can be sufficiently described with a sum,
\be
F^p_{i}(t) &= x_\omega F_{i,\omega}(t) + x_\phi F_{i,\phi}(t) + x_\rho F_{i,\rho}(t), \\
F^n_{i}(t) &= x_\omega F_{i,\omega}(t) + x_\phi F_{i,\phi}(t) - x_\rho F_{i,\rho}(t). 
\label{eq:FFansatz}
\ee
Here we have used $x_\omega = 3(x_u+x_d)$, $x_\phi = -3x_s$ and $x_\rho = (x_u-x_d)$. The corresponding values for each gauge boson model can also be found in \cref{tab:charges}. 

For the electromagnetic force (likewise the dark photon), we have four normalization conditions given by
\be
F_1^p(0)&= 1, \quad &F_1^n(0)&=0, \\ 
F_2^p(0)&=\mu_p - 1,\quad &F_2^n(0)&=\mu_n \,,
\ee
where $\mu_{p,n}$ are the magnetic moments.
In the case of a protophobic model, in which the role of the proton and neutron are exchanged, the conditions should be flipped to
\be
F_1^p(0)&= 0, \quad &F_1^n(0)&=1,\\ 
F_2^p(0)&=\mu_n,\qquad &F_2^n(0)&=\mu_p-1 \ . 
\ee
For the components of the general form factor ansatz of \cref{eq:FFansatz} to fulfill the normalization conditions of both models, this implies
\be
F_{1,\omega}(0) &= 0.5, \quad 
&F_{2,\omega}(0) &= \frac{1}{2}(\mu_p + \mu_n - 1), \\
F_{1,\phi}(0) &= 0,\quad 
&F_{2,\phi}(0) &= 0, \\
F_{1,\rho}(0) &= 0.5, \quad 
&F_{2,\rho}(0) &= \frac{1}{2}(\mu_p - \mu_n - 1) \ .
\ee

As a consistency check, we find at $t=0$,
\be
F_1^p(0) &= x_\omega F_{1,\omega}(0) +x_\phi F_{1,\phi}(0) + x_\rho F_{1,\rho}(0) 
\\ & = 2 x_u \!+\! x_d  \equiv x_p,\\
F_1^n(0) &= x_\omega F_{1,\omega}(0) +x_\phi F_{1,\phi}(0) - x_\rho F_{1,\rho}(0) \\&= x_u + 2 x_d  \equiv x_n \, .
\ee
This result reflects that the fact that the normalization is governed by the valence quark content, and, importantly, that $F_i^p(0)$ is independent of $x_s$, in accordance with the OZI-rule~\cite{Okubo:1963fa,Zweig:1964jf,Iizuka:1966fk}. The latter could equivalently be enforced by directly setting $F_{i,\phi}(0) = 0$. 

\subsection{Resonance Model and Superconvergence}

We model the form factors using a simple vector resonance ansatz with $n$ resonances, expressed as a sum of Breit-Wigner functions,
\be
F_{i,V} = \sum_{j=0}^{n-1} a^{i}_{V_j} {\rm BW}_{V_j}(t) 
\quad \text{for} \quad i=1,2,
\ee
where $V=\omega, \phi, \rho$. The Breit-Wigner function is defined as
\be
{\rm BW}_{V}(t)=\frac{m_{V}^2}{m_{V}^2-t-im_{V}\Gamma_{V}\theta(t)}
\ee
for a resonance state $V$ with mass $m_{V}$ and widths $\Gamma_{V}$ as well as a $\theta$ step function as in Ref.~\cite{Czyz:2014sha}. This parametrization introduces $3\times2\times n=6n$ couplings $a^i_{V_j}$, corresponding to three resonance families and two form factors $F_{1,2}$ which are to be determined either by fitting or through theoretical considerations.

Perturbative QCD constrains the asymptotic behavior of the nucleon form factors at large momentum transfer to be $F_1|_{|t|\to \infty} \sim 1/t^2$ and $F_2|_{|t|\to \infty} \sim 1/t^3$~\cite{Lepage:1980fj}. This condition is expected to hold for any choice of couplings, which in turn implies that the asymptotic scaling must be satisfied separately by each of the individual components $F_{i,V}$. To see the consequences of this requirement, we expand the form factors in inverse powers of $t$:
\be
\!\!F_V(t) 
&= -\frac{1}{t}  \Bigg(\sum_{j=0}^{n-1} a_{V_j}m_{V_j}^2 \Bigg) \\
& \ \ + \frac{1}{t^2} \Bigg(\sum_{j=0}^{n-1} a_{V_j}m_{V_j}^2(m_{V_j}^2\!-\!im_{V_j}\Gamma_{V_j}) \Bigg) \\
& \ \ - \frac{1}{t^3} \Bigg(\sum_{j=0}^{n-1} a_{V_j}m_{V_j}^2(m_{V_j}^2\!-\!im_{V_j}\Gamma_{V_j})^2 \Bigg)  + \dots \, . 
\ee
The desired asymptotic behavior can be enforced by setting the expressions inside the parenthesis to zero.  

For each family of resonances $V$, the form factor must satisfy a set of constraints. For $F_1$, these consist of a normalization condition and a convergence condition enforcing the asymptotic behavior $F_1|_{|t|\to \infty}\to 1/t^2$. For every resonance $V=\rho, \omega, \phi$, these take the form
\be
\sum_{j=0}^{n-1} a_{V_j}^1=F_{1,V}(0),\quad {\rm and} \quad 
\sum_{j=0}^{n-1} a_{V_j}^1 m_{V_j}^2=0 \, , 
\ee
which together provide $3\times 2=6$ independent constraints. For $F_2$, we impose one normalization condition, and two convergence conditions to achieve $F_2|_{|t|\to \infty} \sim 1/t^3$, which respectively are given by
\be
&\sum_{j=0}^{n-1} a_{V_j}^2=F_{2,V}(0), \qquad
\sum_{j=0}^{n-1} a_{V_j}^2 m_{V_j}^2=0,  \\
&\sum_{j=0}^{n-1} a_{V_j}^2 m_{V_j}^2(m_{V_j}^2-im_{V_j}\Gamma_{V_j})=0.
\ee
For vanishing widths $\Gamma_{V_j} =0$, this provides $3\times 3=9$ constraints. With nonzero widths, the last equation must hold separately for imaginary and real parts, giving three additional constraints. 

Let us now consider the case of $n=3$ resonances. For each $F_{1,V}$, we have three parameters $a_{V_i}$ and two conditions. Leaving $a_{V_0}$ as a free parameter, we obtain 
\be
a_1 &= \frac{a_{V_0} m_{V_0}^2 + m_{V_2}^2 [F_{1,V}(0)-a_{V_0}]}{m_{V_2}^2 - m_{V_1}^2},\\
a_2 &= -\frac{a_{V_0} m_{V_0}^2 + m_{V_1}^2[F_{1,V}(0)-a_{V_0}]}{m_{V_2}^2 - m_{V_1}^2}.
\ee
For $F_{2,V}$, we have one normalization constraint, and three convergence conditions (recalling the imaginary part). This system is therefore over-constrained for three real $a_{V_i}$.

One solution is to weaken the requirement on the asymptotic behavior. In Ref.~\cite{Adamuscin:2016rer}, superconvergence was only required for vanishing widths (or equivalently for $t<0$). In that case, the last condition reduces to
\be
\sum_{j=0}^{n-1} a_{V_j}^1 m_{V_j}^4=0 \ .
\ee
In this case, the couplings are uniquely determined as
\be
a_0 &= F_{2,V}(0)\frac{m_{V_1}^2 m_{V_2}^2}{(m_{V_0}^2-m_{V_1}^2)(m_{V_0}^2-m_{V_2}^2)}, \\
a_1 &= F_{2,V}(0)\frac{m_{V_0}^2 m_{V_2}^2}{(m_{V_1}^2-m_{V_0}^2)(m_{V_1}^2-m_{V_2}^2)}, \\
a_2 &= F_{2,V}(0)\frac{m_{V_0}^2 m_{V_1}^2}{(m_{V_2}^2-m_{V_0}^2)(m_{V_2}^2-m_{V_1}^2)}.
\ee
We see that the coefficient vanishes for the $\phi$ case, where $F_{2,\phi}(0)=0$. For $t<0$, the width term in the Breit-Wigner is absent, and the form factor takes the compact form
\be
F_{2,V}^{\Gamma=0}(t) = F_{2,V}(0) \frac{m_{V_0}^2 m_{V_1}^2 m_{V_2}^2}{(m_{V_0}^2-t)(m_{V_1}^2-t)(m_{V_2}^2-t)} \, .
\ee
For $t>0$, however, the full expression is 
\be
&F_{2,V}(t) \!=\! F_{2,V}(0) \,{\rm BW}_{V_0}(t)\,{\rm BW}_{V_1}(t)\,{\rm BW}_{V_2}(t)\\
&\ \ \ \ \ \times\! \Bigg[1 \!+\! \frac{\sum_i {\rm Re(f)}_i  + i \, {\rm Im(f)}_i  }{(m_{r_0}^2-m_{r_{1}}^2)(m_{r_0}^2-m_{r_{2}}^2)(m_{r_1}^2-m_{r_{2}}^2)}  \Bigg],
\ee
where
\be
{\rm Re(f)}_i &= m_im_j\Gamma_i\Gamma_j(m_j^2-m_i^2), \\
{\rm Im(f)}_i &= m_i\Gamma_i(m_j^2-m_k^2)(m_j^2+m_k^2-(m_i^2+t)),
\ee
with $j=(i+1\!\!\mod 3), k=(i+2\!\!\mod 3)$. Note that for $\Gamma_i\to 0$, we have ${\rm Re(f)}_i={\rm Im(f)}_i=0$, recovering the results for $t<0$. We can rearrange the square brackets as $1 + const. + i \, x \cdot t $ where
\be
x \!=\! - \sum_i \frac{m_i \Gamma_i }{(m_{i}^2-m_{j}^2)(m_{i}^2-m_{k}^2)} \ . 
\ee
This sum is dominated by the term proportional to the heaviest mass and largest width, i.e. $m_2 \Gamma_2$. Assuming, for simplicity, that $m_2^2 \gg m_0^2, m_1^2$, one finds $t \cdot x = - \Gamma_2\,   t / m_2^3$. In practice, we have $m/\Gamma > 6$, so this term only becomes significant for $t \gtrsim 6 m_2^2 \gtrsim 20~\gev^2$. In this kinematic regime, however, additional higher resonances and continuum contributions that were ignored in the simple three-resonance model are expected to play a role. Moreover, this region lies outside both the domain of phenomenological interest in this study and the range over which data is fitted.

We have also explored various alternative models. One possibility is a three-resonance model with complex couplings $a_{V_i}$. In this case, the superconvergence constraints can be satisfied exactly even for non-vanishing widths, yielding a closed solution consistent with all conditions. The drawback, however, is that the form factors acquire complex values even for $t<0$. Another possibility is a four-resonance model with real $a_{V_i}$, where the linear system of four equations (one normalization and three convergence conditions) can be solved uniquely for four real couplings.  In practice, however, neither of these alternatives led to substantially different nor improved results compared to the minimal three-resonance realization. We therefore adopt the simplest implementation with the smallest number of free parameters.  

In the three–resonance model, besides fitting the unconstrained coefficients $a_{1,V_0}$, we also allow the masses and widths of the highest-lying resonances to vary. This provides additional flexibility in the fits. To avoid unphysical values, the resonance parameters are varied only within ranges around their PDG values. We then consider three scenarios:  
\begin{itemize}
 \item \texttt{Model-1}: variation of the highest resonances $\omega'', \rho'', \phi''$ around their PDG values,  
 \item \texttt{Model-2}: variation of the two highest resonances $\omega', \omega'', \rho', \rho'', \phi', \phi''$ around their PDG values,  
 \item \texttt{Model-3}: variation of the two highest resonances $\omega', \omega'', \rho', \rho'', \phi', \phi''$ within a wider range around the PDG values.  
\end{itemize}
The details of the fit will be discussed below.

\subsection{Form Factor Fit}

\begin{table*}[tb]
\def\arraystretch{1.3}
\centering
\begin{tabular}{l|l|c}
\hline\hline
Observable & Experiment & $t$ range [GeV$^2$] \\
\hline
\multicolumn{3}{c}{\textbf{Proton}} \\
\hline
$e^+e^-\to p\bar p$ \ \ & BESIII~\cite{BESIII:2015axk}, 
FENICE~\cite{Antonelli:1998fv}, 
DM1~\cite{Delcourt:1979ed}, 
CLEO~\cite{CLEO:2005tiu}, 
ADONE~\cite{Castellano:1973wh} & [3.6,13.5] \\
$p\bar p\to e^+e^-$ & E760~\cite{E760:1992rvj}, 
E835~\cite{E835:1999mlt,Andreotti:2003bt} & [8.8,14.4] \\
$|G_E^p/G_M^p|$ & BABAR~\cite{BaBar:2013ves} & [3.7,7.3] \\
$\mu_p G_E^p/G_M^p$ & JLab~\cite{Puckett:2011xg,Puckett:2017flj,Punjabi:2005wq} & [–8.5,-1.2] \\
$G_E^p$ & E94110~\cite{E94110:2004lsx}, 
Bernauer~\cite{Bernauer:2010zga}, 
BESIII~\cite{BESIII:2019hdp}, others & [–0.015,9.5] \\
$G_M^p$ & Hohler~\cite{Hohler:1976ax}, 
Bernauer~\cite{Bernauer:2010zga}, 
DESY~\cite{Bartel:1973rf}, 
Mark~III~\cite{Janssens:1965kd}, 
BESIII~\cite{BESIII:2019hdp}, 
E835~\cite{E835:1999mlt}, others & [–0.015,14.4] \\
$G_{\rm eff}^p$ & BABAR~\cite{BaBar:2013ves}, 
BESIII~\cite{BESIII:2019hdp,BESIII:2015axk,BESIII:2021rqk} & [3.6,17.0] \\
\hline
\multicolumn{3}{c}{\textbf{Neutron}} \\
\hline
$G_E^n/G_M^n$ & JLab~\cite{JeffersonLaboratoryE93-038:2005ryd}, 
BLAST~\cite{BLAST:2008bub} & [–1.47,-0.14] \\
$G_E^n$ & Golak~\cite{Golak:2000nt}, 
A1~\cite{Glazier:2004ny}, 
JLab~\cite{JeffersonLabE93-026:2003tty,E93026:2001css}, 
MAMI~\cite{Herberg:1999ud,Ostrick:1999xa}, NIKHEF~\cite{Passchier:1999cj}, OHIPS~\cite{Eden:1994ji}, Berard~\cite{Berard:1973xrv} & [–1.0,-0.002] \\
$G_M^n$ & SLAC~\cite{Rock:1982gf,Lung:1992bu}, 
JLab~\cite{JeffersonLabE95-001:2006dax}, 
MAMI~\cite{Kubon:2001rj}, 
Mainz-A1~\cite{Anklin:1998ae} & [–10.0,-0.07] \\
$G_{\rm eff}^n$ & BESIII~\cite{BESIII:2021tbq}, 
SND~\cite{Druzhinin:2020fgn} & [3.5,9.5] \\
\hline\hline
\end{tabular}
\caption{Summary of proton and neutron form factor data used in this work. The proton dataset contains $\sim$ 465 points in the range $t\in[-8.5,17]$~GeV$^2$, while the neutron dataset contains $\sim$ 82 points in the range $t\in[-10,9.5]$~GeV$^2$.}
\label{tab:ff_data}
\end{table*}

The fit relies on a broad set of experimental measurements of nucleon form factors in both the space-like $t<0$ and time-like $t>0$ regions. They are summarized in Tab.~\ref{tab:ff_data} for proton and neutron data. Experimental data is usually given in terms of the Sachs form factors,
\be
    G_E^N(t)&=F^N_1(t)- \tau F^N_2(t),\\
    G_M^N(t)&=F^N_1(t)+F^N_2(t),
\ee
for $N = p,n$ with $\tau=-t/(4m_N^2)$ and the effective form factor
\be
    G_{\rm eff}^N\equiv \sqrt{\frac{|G_E^N|^2+\frac{t}{2m_N^2}|G_M^N|^2}{1+\frac{t}{2m_N^2}}}.
\ee
For the proton, we include cross section data from $e^+e^-\to p\bar{p}$ and $p\bar{p}\to e^+e^-$ as well as measurements of $G_E^p, G_M^p$, $|G_E^p/G_M^p|$ and the effective form factor $G_{\rm eff}^p$ from a variety of fixed-target and collider experiments. In addition, precise polarization measurements from JLab provide constraints on the ratio $\mu_p G_E^p/G_M^p$ over a wider range of $t$ values. For the neutron, we use data on $G_E^n$, $G_M^n$ and $G_E^n/G_M^n$ from scattering and polarization experiments, as well as recent time-like measurements from BESIII and SND. 

Together, these datasets cover an energy range $-10~\gev^2 < t < 17~\gev^2$ from the very low-$t$ regime dominated by static properties up to higher momentum transfers where we expect the perturbative behavior to set in. 

However, there remains a so-called `unphysical region` between $0<t<4m_p^2$, i.e. below the proton-antiproton production threshold, where no direct experimental data is available. In this kinematic window, one must rely entirely on theoretical constraints. These include the normalization conditions at $t=0$, which fix the static charges and magnetic moments, and, as discussed earlier, reflect that the normalization is governed by the valence quark content. In addition, the $\phi$ contribution to $F_2$ vanishes in accordance with the OZI rule, and the form factor decomposition in this region reflects the physical meson content of the nucleon current. The associated model uncertainty can be assessed by employing several resonance parameterizations and estimating the spread among their predictions. 

One of the central challenges of our fit arises from the imbalance between the large set of experimental data, in part extremely precise, and the relatively small number of free parameters in our model. As a consequence, it is not surprising that the parametrization does not provide an excellent description of all observables simultaneously. More importantly, when using the standard definition of the $\chi^2$ function, the extracted parameter uncertainties turn out to be unrealistically small. This indicates that the naive $\chi^2$ approach does not yield a meaningful estimate of the true model uncertainty. 

To overcome this problem, we introduce an additional model uncertainty $\sigma_{\text{model}}$, which regularizes the fit by effectively de-weighting overly precise data points~\cite{Hogg:2010xxx}. The modified $\chi^2$ function is defined as
\be
\!\!\! \chi^2 \!=\! \sum_{i} \frac{(y_{model,i}\!-\!y_{data,i})^2}{\sigma_{data,i}^2 + \sigma_{model,i}^2} + \log\Bigg(\!1\!+\!\frac{\sigma_{model,i}^2}{\sigma_{data,i}^2}\!\Bigg) .
\label{eq:chi2}
\ee
For $\sigma_{model}=0$, this reduces to the usual $\chi^2 = \sum_{i} (y_{model,i}-y_{data,i})^2 / \sigma_{data,i}^2$. However, for $\sigma_{model}>0$ the contribution of the most precise data points is reduced, preventing them from dominating the fit. The logarithmic terms were introduced to penalize large model uncertainties. In practice, we adopt the simple ansatz $\sigma_{model} = f y_{model}$ with a constant fractional uncertainty $f$ that is common to all observables. 

In addition to the model uncertainty described above, we further regularize the fit by including two penalty terms in the $\chi^2$ function. First, as discussed earlier, we allow the masses and widths of the highest resonances, $m_{V_i}$ and $\Gamma_{V_i}$ with $V=\omega,\rho,\phi$, to vary within the PDG uncertainties~\cite{ParticleDataGroup:2024cfk}. To ensure that these variations remain within the PDG uncertainties, we add the term
\be
\chi^2_{\rm add,1} =\sum_{V_i} \frac{(m_{V_i} - m_{V_i}^\text{PDG})^2 }{(\Delta m_{V_i}^\text{PDG})^2} + \frac{(\Gamma_{V_i} - \Gamma_{V_i}^\text{PDG})^2}{(\Delta \Gamma_{V_i}^\text{PDG})^2}
\ee
to the $\chi^2$ function for varying the higher resonance $V_i$. 

Second, we impose a mild condition that disfavors large absolute values of the resonance coefficients $a_{V_i}$. This is motivated by the expectation that the physical couplings should not be unnaturally large, avoiding dominance by a single meson and strong cancellations between different mesons, and preventing the fit from settling into false minima. The corresponding penalty term is chosen as
\be
\chi^2_{\rm add,2}=\sum_i |a_{V_i}|^2.
\ee
Together, these additional contributions keep the parameters within physically reasonable ranges while preserving sufficient flexibility to describe the data. 

For the fit procedure, we employed both \texttt{scipy}'s optimization routines~\cite{2020SciPy-NMeth} and the \texttt{iminuit} package~\cite{James:1975dr, iminuit}, obtaining consistent results between the two. We observed that the choice of initial parameter values could influence convergence and the final miniumum. To mitigate this dependence, we generated 500 random initial parameter sets, performed fits from each, and selected the configuration yielding the lowest $\chi^2$. For this final minimization, we adopted \texttt{iminuit}, which is particularly efficient for large datasets and offers robust error estimation.

Different weighting schemes were also explored to account for the varying precision and abundance of individual datasets. However, with the inclusion of the model uncertainty term described in \cref{eq:chi2}, equal weights across all datasets were sufficient. The additional uncertainty effectively de-weights overly precise or abundant datasets, preventing them from dominating the fit and yielding more balanced parameter estimates.

\begin{figure*}[t]
\centering
\includegraphics[width=0.95\textwidth]{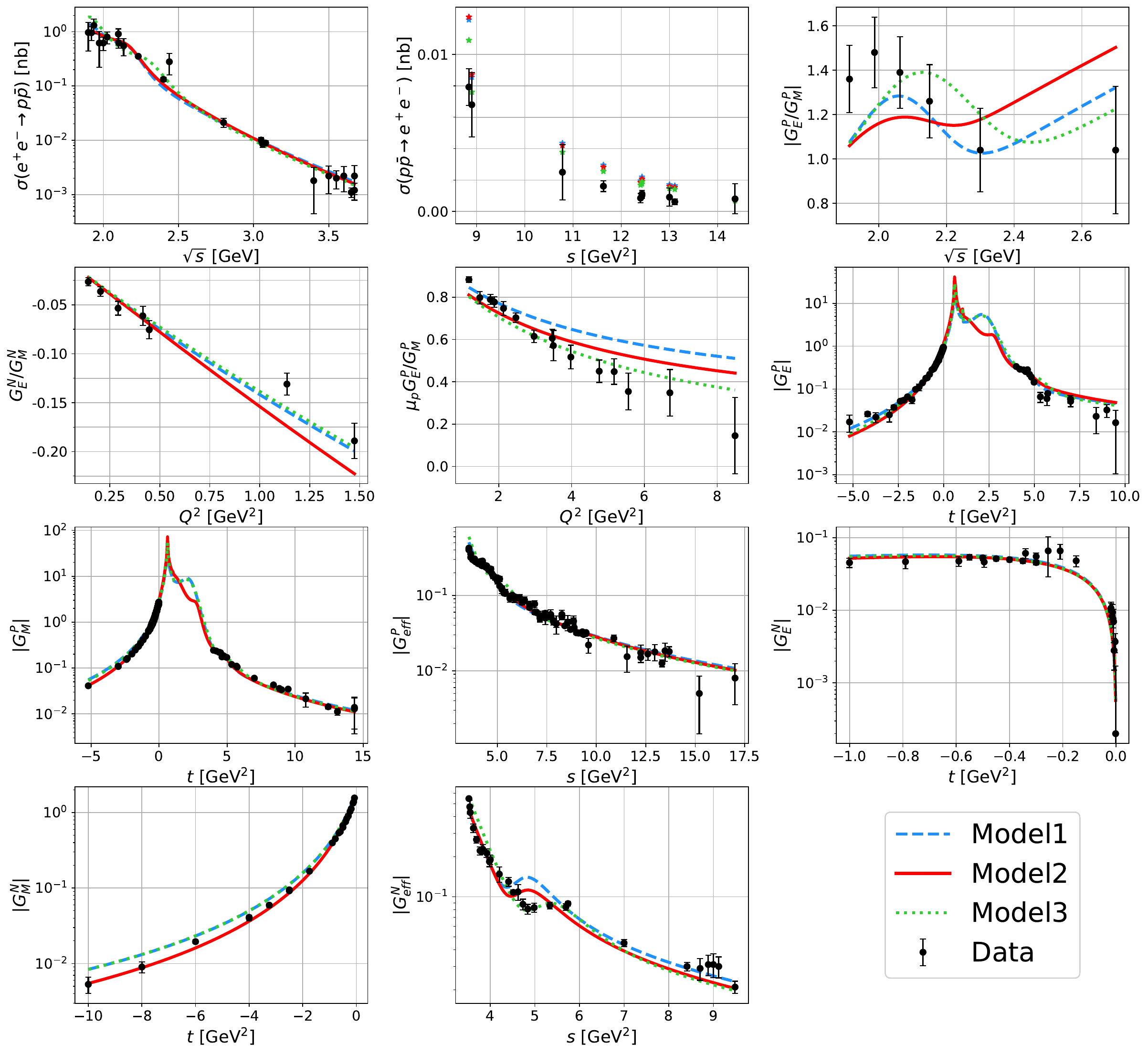} 
\caption{Comparison of the fitted model predictions with the 11 observables included in the global fit. Each panel shows the measured data (black points with error bars), and the best-fit model prediction (colored lines). The details of the models are described in \cref{sec:ffs}, whereas the corresponding data can be found in \cref{tab:ff_data}. For panels containing exclusively timelike data, the horizontal axis shows $\sqrt{s}$ or $s$; for panels with exclusively spacelike data, the horizontal axis shows $Q^2=-t$ or $t$; and for panels containing both timelike and spacelike data, the horizontal axis is labeled with $t$.}
\label{fig:fit}
\end{figure*}

The results of the fit are shown in \cref{fig:fit}. All fitted models show good agreement with the data from the 11 observables, with some models performing slightly better for certain observables than others. Overall, however, the agreement is close, which represents a remarkable achievement given the simplicity of the paramtrization. In \cref{fig:fit}, \texttt{Model-1} is shown as blue dashed, \texttt{Model-2} as red solid, and \texttt{Model-3} as a green dotted line. Throughout the fitting, ratios were generally more difficult to handle; in particular, the proton ratio $|G_E^P/G_M^P|$, as well as $\mu_pG_E^P/G_M^P$, were challenging, due to the relatively limited amount of data available compared to other datasets to constrain the fit. For \texttt{Model-3}, allowing a wider range around the PDG values of higher resonance mass and width values proved beneficial for $G_{\rm eff}^N$, enabling the curve to accommodate a resonance-like feature around $s\sim 5-6$~GeV$^2$. In all three cases, the model uncertainty $\sigma_{model}$ was around 10\% and hence smaller than the difference between the models. We therefore used the spread of the three models as a measure of uncertainty.

\begin{figure*}[t]
\centering
\includegraphics[width=\textwidth]{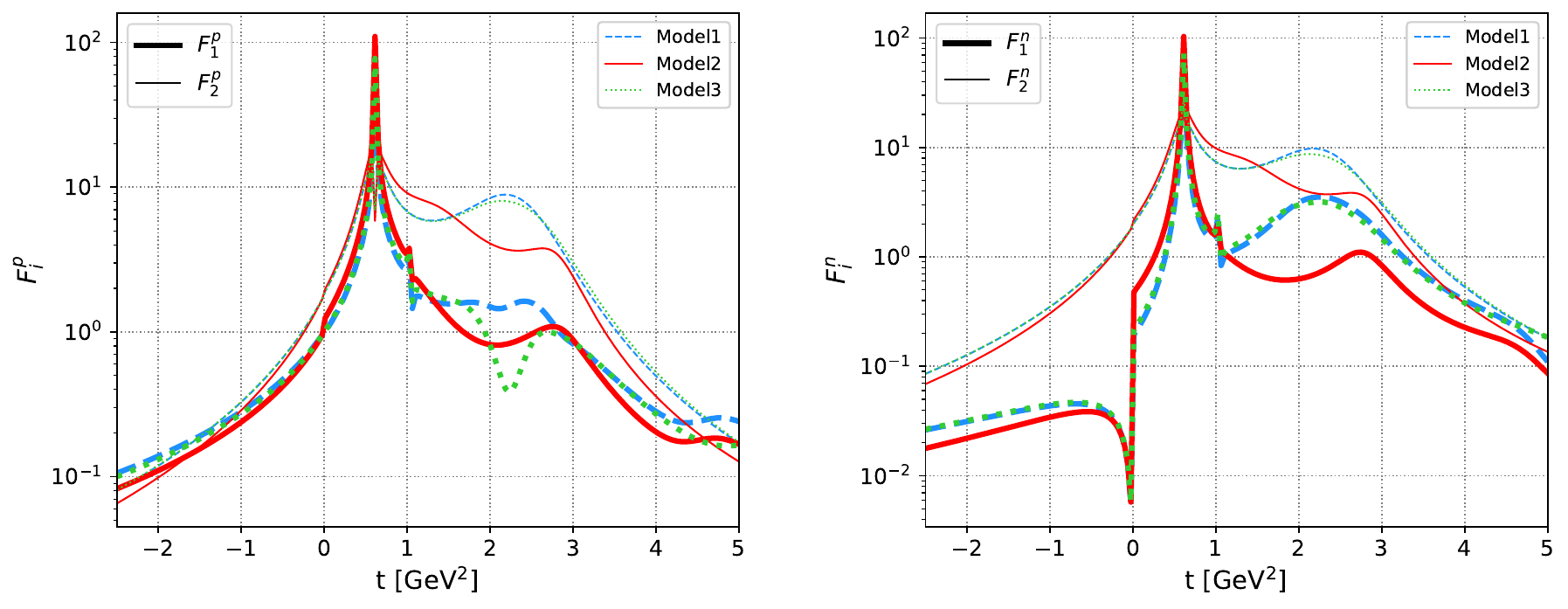} 
\caption{Form factors $F_1$ (thick lines) and $F_2$  (thin lines) for all form factor models mentioned in \cref{sec:ffs} for the proton (\textbf{left}) and neutron (\textbf{right}). We show the electromagnetic form factors for the SM, but the exact same form factors can be used the dark photon since $x_{\omega,\rho,\phi}=1$ as seen from \cref{tab:charges}.}
\label{fig:F1F2}
\end{figure*}

The final form factor plots are shown in \cref{fig:F1F2}, where we present the SM case for the $F_1$ (thin lines) and $F_2$ (thick lines) form factors for the proton and neutron for all models discussed in \cref{sec:ffs}. The largest variation between the form factor models occurs in the intermediate range between $t\sim 1-3$~GeV$^2$ where higher resonances dominate. This behavior is expected, as these resonances were allowed to vary in the fit and their masses and widths are experimentally less well constrained than those of the lowest-lying states. At $t=0$~GeV$^2$, all models are constrained by the normalization condition, while at large $t$ they follow the power-law falloff consistent with perturbative QCD. The differences in the intermediate region therefore represent the main source of model dependence in our predictions.

Since for the dark photon case, all individual meson contributions enter proportionally to those of the SM photon, so $x_{\omega, \phi, \rho}=1$,  the same form factors can be applied directly. For all other models, the form factors are modified according to \cref{eq:FFansatz} with the charges given in \cref{tab:charges}.

\section{Modeling of Bremsstrahlung}
\label{sec:modeling}

\begin{figure*}[t]
\centering
\includegraphics[width=0.49\textwidth]{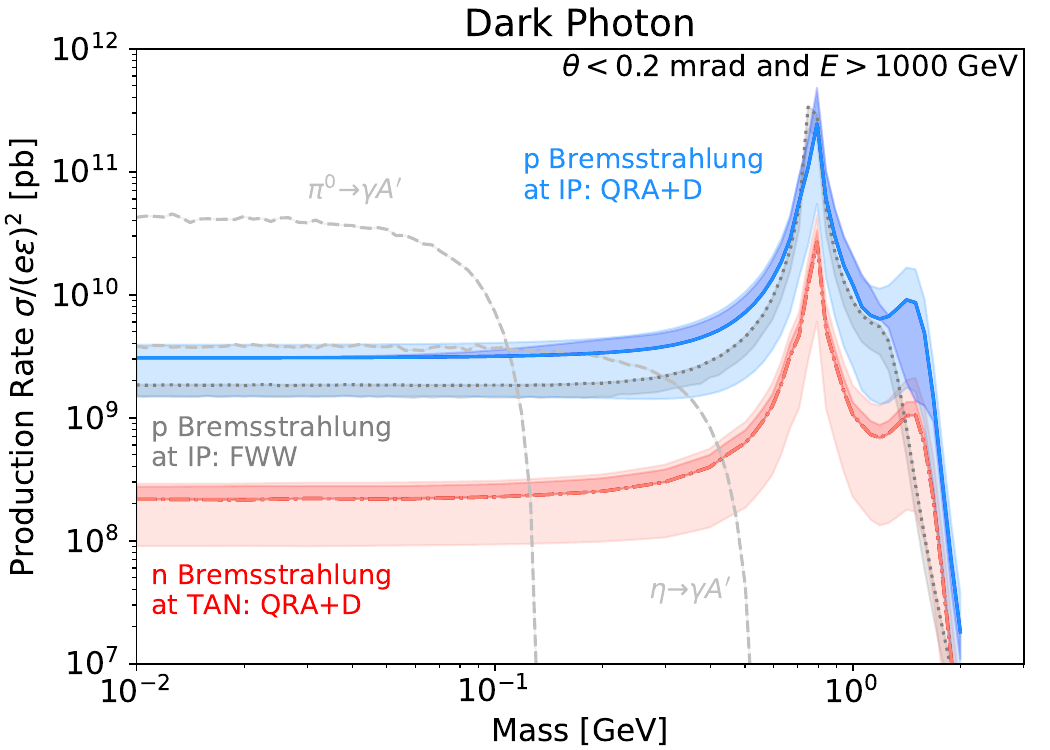} 
\includegraphics[width=0.49\textwidth]{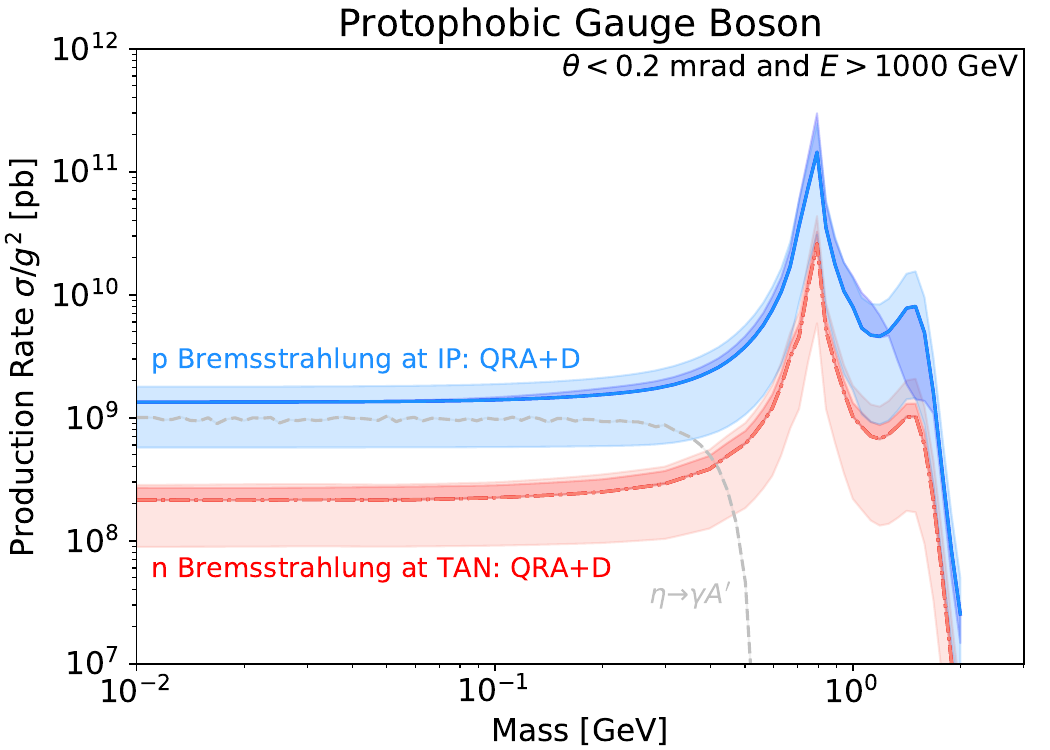} 
\caption{The cross section in FASER's signal region ($\theta<0.2$~mrad and $E>$TeV), for dark photons (\textbf{left}) and protophobic gauge bosons (\textbf{right}). The predicted rate from proton bremsstrahlung at the IP and neutron bremsstrahlung occurring in the TAN are shown in blue and red, respectively. The inner darker shaded regions correspond to a variation of the form factor model as well as the neutron production model. The lighter outer bands correspond to additionally varying the value of $\Lambda_p$ from 1 to 2~GeV. For the Dark photon case, we show the previously used FWW approximation prediction as gray dotted curve. Shown as gray dashed lines are the production rates from meson decays.}
\label{fig:prod_vs_mass}
\end{figure*}

Let us now turn to the modeling of light gauge boson production via Bremsstrahlung. The double-differential dark-photon production cross section can be written in the factorized form~\cite{Feng:2017uoz, Foroughi-Abari:2021zbm},
\be
\!\!\! \frac{d\sigma}{dp d\theta} =  (g/e)^2 \,  J \, \sigma^{nsd}_{pp}(s') \, w(z,p_T^2) \, |F(m_{A'}^2)|^2 \, V(p_{A'}),
\label{eq:dsig}
\ee
where $p$ and $\theta$ are the dark-photon momentum and polar angle with respect to the beam, and $g$ is the gauge coupling. We further use the fraction of the beam momentum carried by the dark photon $z\approx p/E_{\rm beam}$ and the dark photon transverse momentum $p_T\approx p\cdot \theta$ as kinematic variables. The Jacobian determinant $J$, corresponding to the variable transformation from $(z,p_T^2)$ to $(p,\theta)$ is given by
\be
J \;=\; \left|\det\frac{\partial(z,p_T^2)}{\partial(p,\theta)}\right|
\approx 2 p_T z .
\ee
$\sigma^{nsd}_{pp}(s')$ is the cross section evaluated at a reduced center of mass energy $s' \approx (1-z)s$. Here we follow the argument of Ref.~\cite{Foroughi-Abari:2021zbm} that initial and final state radiation amplitudes for elastic and single diffractive scattering interfere destructively and use only the non-single diffractive cross section from Ref.~\cite{Likhoded:2010pc} in which no such cancellations occur. 

For the splitting functions $w$ we adopt the splitting kernel derived via the quasi-real approximation including the Dawson correction (QRA+D) as presented in Ref.~\cite{Foroughi-Abari:2024xlj}. Unlike older descriptions, this result accounts for contributions for the Dirac form factor $F_1$ and Pauli form factor $F_2$. The splitting part of \cref{eq:dsig} becomes
\be
w\,|F|^2 &\longrightarrow
w^{\rm QRA+D}(z,p_T^2)\,|F_1(m_{A'}^2|^2 \\
& \ \ \ + w_2^{\rm QRA+D}(z,p_T^2)\,|F_2(m_{A'}^2)|^2\\
&  \ \ \ + w_{12}^{\rm QRA+D}(z,p_T^2)\,{\rm Re}\big(F_1(m_{A'}^2) F_2^*(m_{A'}^2)\big) ,
\label{eq:w_F1F2}
\ee
where the form factors are evaluated at $t=m_{A'}^2$.

The factor $V(p_{A'})$ accounts for the various approximations and  validity conditions required in the QRA+D derivation. In particular, we require co-linearity, meaning that with $z \simeq E_{A'}/E_{\rm beam}$ one must have $z \gg \text{max}(m_p,\, p_T,\, m_{A'})/E_{\rm beam}$. In addition, we demand $H / (4E_p^2 z (1-z)^2) < 1 $ where $H = p_T^2 + (1-z)\,m_{A'}^2 +z^2\,m_p^2$. For the kinematic region of interest for FASER, these condition are automatically satisfied.

An additional constraint arises from the validity of the Bremsstrahlung picture. The incoming proton radiates off a dark photon and turns into an off-shell proton with four-momentum $p'$ which then undergoes a scattering with the target nucleus. This approximation assumes that the intermediate off-shell proton stays intact and does not break up. To limit contributions from highly off-shell recoiling protons, which would lead to break-up, we include the damping factor suggested in Ref.~\cite{Foroughi-Abari:2021zbm},
\be
f(p'^2)=\frac{\Lambda_p^4}{\Lambda_p^4 + (p'^2-m_p^2)^2},
\ee
where $p'^2$ can be approximated as $ p'^2 \simeq m_p^2 - H / z$. The canonical choice is given by $\Lambda_p = 2 \pi \Lambda_\text{QCD}\sim 1.5~\gev$ and the variation of $1.0~\gev $ to $2.0~\gev$ is used to estimate the uncertainty~\cite{Foroughi-Abari:2024xlj}. 

Before proceeding to examples, we stress again that the Bremsstrahlung formalism and the derived form factors apply to models of dark vector bosons with arbitrary couplings and for both proton and neutron Bremsstrahlung. In all cases, only the form factors $F_1$ and $F_2$ change, as described in \cref{eq:FFansatz}.

\section{Improved Predictions for FASER}
\label{sec:FASER}

Having discussed our formalism to model Bremsstrahlung, let us now turn to its application. As an example, we consider the FASER experiment and recast their results using our improved predictions, first for dark photons in this section and then for other models in the next section.

\subsection{Experimental Setup}

FASER is a far-forward experiment located about 480~m downstream of the ATLAS interaction point (IP) in a previously unused side tunnel of the LHC~\cite{FASER:2022hcn}. It is aligned with the beam collision axis and covers polar angles $\theta \lesssim 0.2$~mrad or equivalently pseudorapidities $\eta \gtrsim 9$. Located at its front is the FASER$\nu$ neutrino detector to observe and study neutrinos at the LHC~\cite{FASER:2019dxq, FASER:2020gpr, FASER:2025qaf, FASER:2024hoe}. Located behind is FASER's long-lived particle detector~\cite{FASER:2018ceo, FASER:2018bac, FASER:2018eoc, FASER:2019aik, FASER:2021ljd}. It consists of a cylindrical decay volume with 1.5~m length and 10~cm radius, which is preceded by a veto system and followed by a spectrometer, a pre-shower and an electromagnetic calorimeter. 

Thus far, FASER has performed two searches for long-lived particles. The first search, published in 2023, targeted dark photons decaying to electrons~\cite{FASER:2023tle}. It required the front veto scintillators to stay inactive, the presence of two tracks, and a calorimeter energy of more than 500~GeV. No event was seen in a dataset corresponding to an integrated luminosity of $27~\ifb$, and constraints were set on the dark photon and $U(1)_{B-L}$ gauge boson models. The second search published in 2024 targeted axion-like particles decaying to photons~\cite{FASER:2024bbl}. Signal events were required to have no activity in the front veto scintillators, a pre-shower signal consistent with the start of an electromagnetic shower, and an energy deposit of more than 1.5~TeV in the calorimeter. Since no tracking requirements were made, this search utilizes both the combined decay and spectrometer volume as fiducial volume, which is roughly described by a cylinder with radius of 10~cm and length of 4~m. One event was found in a dataset corresponding to an integrated luminosity of $65~\ifb$, consistent with the background expectation from neutrinos that interacted in the pre-shower.

While the analysis was designed to search for long-lived particles decaying into multi-photon final states, the analysis was also sensitive to decays to electrons. Constraints derived from this search on dark photons were presented by the FASER collaboration. This interpretation used the Fermi-Weizs\"acker-Williams (FWW) approximation, as described in Refs.~\cite{Blumlein:2013cua, Feng:2017uoz}, to model dark photon Bremsstrahlung. Model uncertainties were estimated by varying an upper transverse momentum cut by a factor of two around its nominal value of 1~GeV. In the following, we reinterpret these results using the updated modeling of Bremsstrahlung using the QRA+D splitting functions and the form factors derived above. 

In the left panel of \cref{fig:prod_vs_mass}, we show the production cross section of dark photons at the LHC as a function of their mass. Here we applied selection cuts $\theta < 0.2~\mrad$ and $E > 1~\tev$ which roughly correspond to the region probed by FASER. The blue solid line shows the production cross section via proton Bremsstrahlung occurring in the primary interactions at the IP. The central prediction corresponds to \texttt{Model-1} with $\Lambda_p=1.5~\gev$. The inner band (dark blue shaded) shows how this prediction changes when varying the form factor model. The outer band (light blue shaded) corresponds to variations of $\Lambda_p$ by 0.5~GeV. We can see that the latter gives larger uncertainties.  

The gray dotted line shows the prediction of the FWW approximation. The overall rate is consistent with the QRA+D prediction within uncertainties, except for the highest masses $m_{A'} > 1.5~\gev$. The gray dashed lines also show the production rate of dark photons with neutral pion and $\eta$ meson decays: while pion decays dominate the overall rate when kinematically open, $\eta$-meson decays contribute similarly to proton Bremsstrahlung. We note, however, that the energy spectra are different, which further changes the relative importance. 

\subsection{Neutron Bremsstrahlung in the TAN}

\begin{figure*}[t]
\centering
\includegraphics[width=0.49\textwidth]{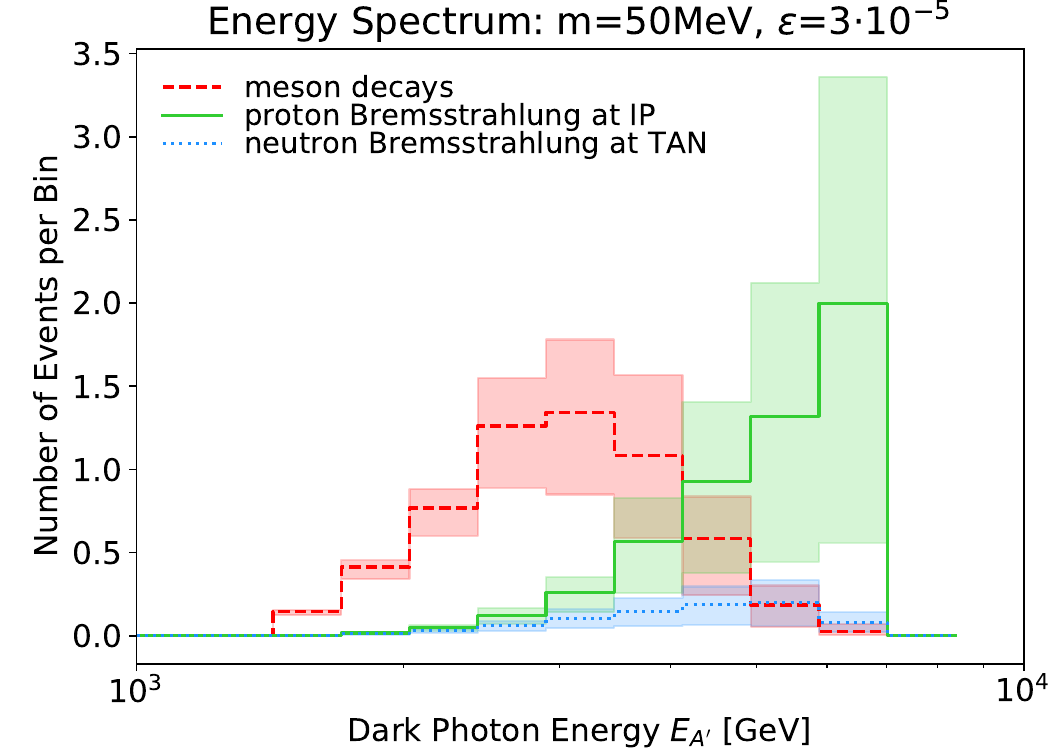} 
\includegraphics[width=0.49\textwidth]{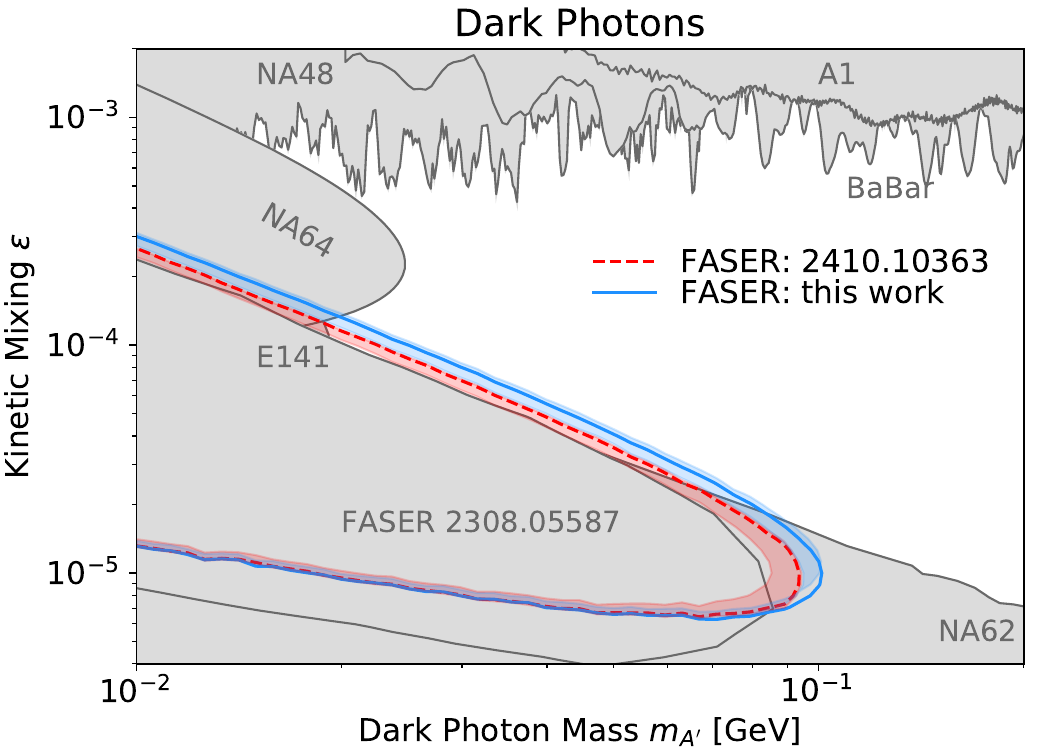} 
\caption{\textbf{Left:} Expected energy spectrum of a dark photon with $m_{A'}=50$~MeV and $\epsilon=3\cdot 10^5$ at FASER. Contributions from meson decays, proton Bremsstrahlung at the IP and neutron Bremsstrahlung at the TAN are shown as red dashed, green solid and blue dotted curves, respectively. The shaded regions show the uncertainty band. \textbf{Right:} Recasted sensitivity of FASER's search on ALPs presented in Ref.~\cite{FASER:2024bbl} on the dark photon parameter space. The red dashed curve corresponds to the bounds presented by the FASER collaboration, using the FWW approximation to simulate proton Bremsstrahlung. The blue curve shows our recasted result using the QRA+D splitting functions  and the new form factors for both proton Bremsstrahlung at the IP and neutron Bremsstrahlung at the TAN. The gray shaded region correspond to previous bounds.}
\label{fig:DarkPhoton}
\end{figure*}

In addition to proton Bremsstrahlung at the primary collision, dark photons may also be produced via Bremsstrahlung of neutrons. This may seem counter-intuitive, since dark photons couple proportional to charge and neutrons are electrically neutral. We note, however, that the form factor $F_1(t)$ only vanishes for $t=0$ while dark photon production is sensitive to $t=m_{A'}^2$. In addition, the dark photon interacts via the Pauli form factor $F_2(t)$ which does not vanish for any value of $t$. 

The LHC produces a large flux of neutrons with multi-TeV energies in the forward direction, which has been measured by LHCf~\cite{LHCf:2018gbv}. At the highest rapidities, these neutrons will be absorbed by the TAN, which is a neutral particle absorber placed about 150~m downstream from the interaction point where the beam pipe splits into separate pipes for the two beams. Dark photons originating in neutron Bremsstrahlung events at the TAN will have to travel a shorter distance than those produced at the primary collision in ATLAS, allowing to push FASER's sensitivity to smaller lifetimes and hence larger couplings.   

To obtain the flux of dark photons we simulate their production in neutron fixed target collisions by numerically integrating over the possible neutron energies and rapidities. In particular, we use the Monte Carlo generators \texttt{EPOSLHC}~\cite{Pierog:2013ria}, \texttt{SIBYLL}~2.3d~\cite{Riehn:2019jet} and \texttt{QGSJET~2.04}~\cite{Ostapchenko:2010vb} to model neutron production and its uncertainty. The resulting dark photon production rate within FASER's simplified acceptance of $\theta<0.2$~mrad and $E>1$~TeV as a function of dark photon mass is shown in \cref{fig:prod_vs_mass} as indicated by the red dash-dotted line. The central prediction was obtained using \texttt{EPOSLHC} for neutron production, form-factor \texttt{Model-1} and $\Lambda_p=1.5$~GeV. The inner (darker) shaded regions corresponds to the uncertainty obtained by varying the form factor model as well as the neutron production model. The outer (lighter) shaded band corresponds to varying  $\Lambda_p$ from 1 to 2~GeV. As before, we can see that the variation of  $\Lambda_p$ provides the leading source of uncertainty. 

\subsection{Dark Photon Results}

We are now ready to look at the dark photon spectrum and sensitivity at FASER. As mentioned before, we consider the setup corresponding to the 2024 analysis with an integrated luminosity of $65~\ifb$. The signal yield is modeled using the \texttt{FORESEE} package~\cite{Kling:2021fwx} assuming a half crossing angle of 160~mrad and the energy dependent signal efficiencies provided by the FASER collaboration in Ref.~\cite{FASER:2024bbl}. 

The left panel of \cref{fig:DarkPhoton} shows the energy spectrum of dark photons decaying in FASER after accounting for efficiencies for $m_{A'}=50~\mev$ and $\epsilon=3\cdot 10^{-5}$. Dark photons from neutral pion and $\eta$-meson decays are shown as a red dashed line, with a modeling uncertainty defined by the spread of Monte-Carlo generators.  The contributions to dark photons from proton Bremsstrahlung in the primary collisions interaction point and neutron Bremsstrahlung occurring at the TAN are shown as green solid line and blue dotted line, respectively. The corresponding uncertainties are defined by the variation of form factor models, $\Lambda_p$ and the neutron production model, where we remind the reader that the uncertainty from $\Lambda_p$ variation dominates. We can see that proton Bremsstrahlung dominates over meson decays at the highest energies, while neutron Bremsstrahlung is sub-dominant. 

Let us now turn to the updated sensitivity of the search presented in Ref.~\cite{FASER:2024bbl}. Limits are set on models predicting more than 3.6~events, which corresponds to the 90\%~C.L exclusion region considering the expected background of $0.44 \pm 0.39$ events and one observed event as estimated via the \texttt{pyhf} package~\cite{Heinrich:2021gyp}. The resulting sensitivity in the dark photon parameter space, spanned by its mass $m_{A'}$ and coupling $\epsilon$, is shown in the right panel of  \cref{fig:DarkPhoton}. The gray shaded regions are those previously excluded by searches at E141~\cite{Andreas:2012mt}, A1~\cite{Merkel:2014avp}, BaBar~\cite{BaBar:2014zli}, NA48~\cite{NA482:2015wmo}, NA64~\cite{NA64:2019auh}, NA62~\cite{NA62:2023nhs}, and FASER~\cite{FASER:2023tle}. 

The red dashed line and uncertainty band show the sensitivity as presented in Ref.~\cite{FASER:2024bbl}, where the FWW approximation was used to model dark photon production via Bremsstrahlung. We have confirmed that we can reproduce this result using our simulation framework. The solid blue line and uncertainty band show the sensitivity obtained when using the QRA+D splitting functions and updated form factors to model dark photon production in proton and neutron Bremsstrahlung. The reach increases noticeably compared to the official published bounds.

\section{Application to other models}
\label{sec:application}

\begin{figure*}[t]
\centering
\includegraphics[width=0.49\textwidth]{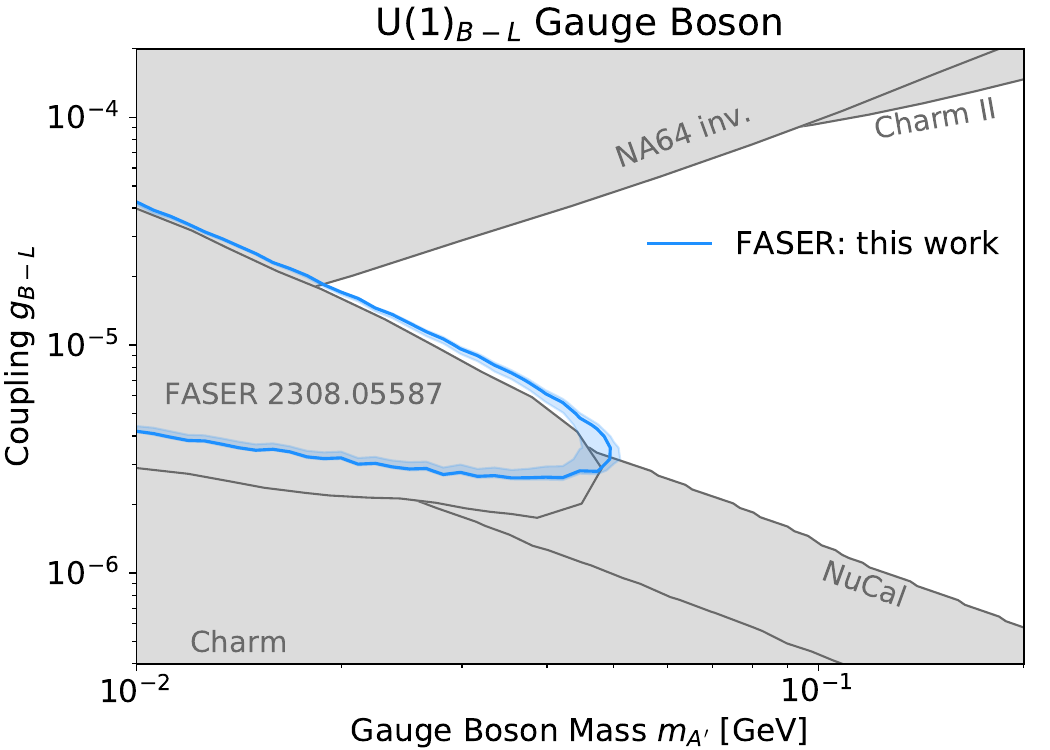} 
\includegraphics[width=0.49\textwidth]{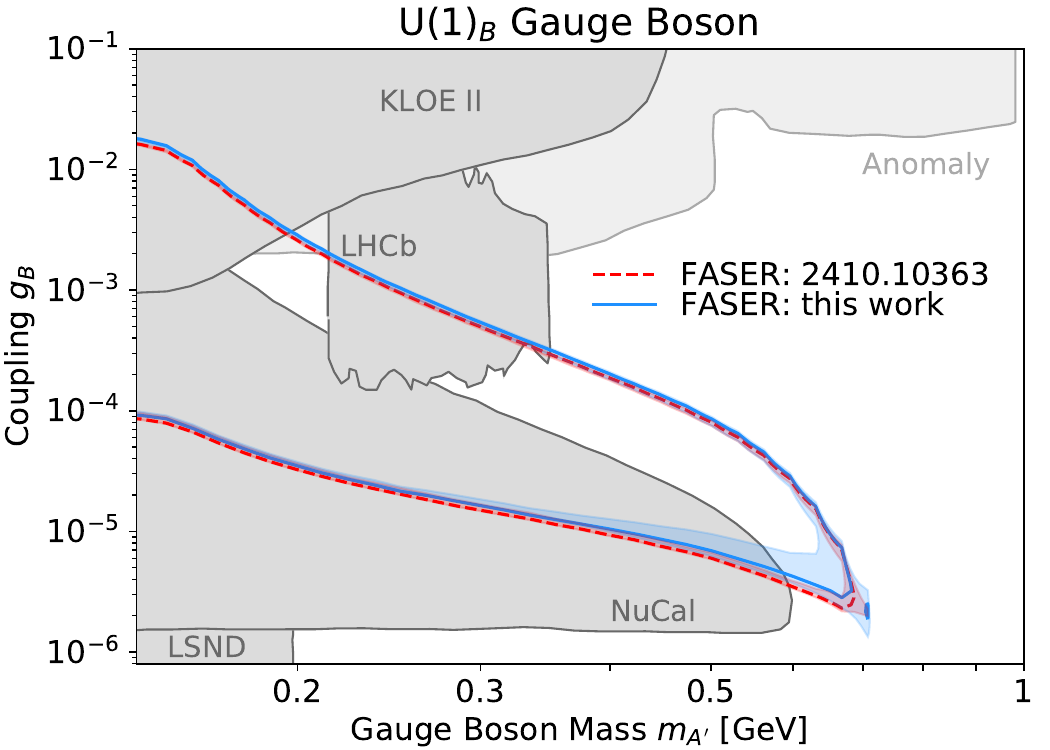} 
\caption{Recasted sensitivity of FASER's ALP search in the paramater space of a U(1)$_{B-L}$ gauge boson (\textbf{left}) and a U(1)$_{B}$  gauge boson (\textbf{right}). The red curve corresponds to the bounds presented by the FASER collaboration in Ref.~\cite{FASER:2024bbl}, using the FWW approximation to simulate proton Bremsstrahlung. The blue curve shows our result using the QRA+D splitting functions and new form factors for both proton Bremsstrahlung at the IP and neutron Bremsstrahlung at the TAN. The gray shaded region correspond to previous bounds.}
\label{fig:U1B}
\end{figure*}

The form factors derived in \cref{sec:ffs} were specifically designed such that they can also be applied to model the production of other light gauge bosons with different coupling structures. In the following, we apply our method to the gauge bosons of the U(1)$_{B-L}$, U(1)$_{B}$ and protophobic gauge groups and reinterpret the results of FASER's ALP search for these models. We further show how our results can be used to model millicharged particle production in proton Bremsstrahlung. 

\subsection{U(1)$_{B-L}$ and U(1)$_{B}$ Gauge Boson}

Both baryon number $B$ and total lepton number $L$ are globally conserved within the SM. This motivates the idea that such conserved quantities might originate from local gauge symmetries. A particularly well-motivated case is the $U(1)_{B-L}$ gauge symmetry, which is conserved at the classical level and, once three sterile neutrinos are introduced to generate neutrino masses, remains free of quantum anomalies. Another frequently studied possibility is a $U(1)_{B}$ gauge symmetry. Unlike $U(1)_{B-L}$, however, $U(1)_{B}$ is anomalous within the SM and therefore requires additional new physics at higher energy scales to achieve anomaly cancellation. Both scenarios predict the existence of a corresponding gauge boson, analogous to the dark photon. Such a boson could be light, with masses in the MeV–GeV range, and feebly coupled ($g \sim 10^{-5}$), leading to a long lifetime.
Both models share the same coupling structure to quarks, as summarized in \cref{tab:charges}. The most significant difference from dark photons is that these gauge bosons do not mix with the $\rho$, but only with the $\omega$ and $\phi$. This results in a modified form factor and a different production cross section via Bremsstrahlung, which we simulate in close analogy to the dark photon case. For the signal yield predictions, we use the decay widths and lifetimes provided by the \texttt{DeLiVeR} package~\cite{Foguel:2022ppx}. At the masses of interest, the $U(1)_{B-L}$ gauge boson exclusively decays to electron pairs, while the $U(1)_{B}$ gauge boson primarily decays to a $\pi^0 \gamma$ final state. 

In the left and right panel of \cref{fig:U1B} we present the parameter space of the $U(1)_{B-L}$ and $U(1)_{B}$ gauge boson, spanned by their mass and gauge coupling, respectively.  Shown in gray are previously excluded regions. The bounds for the $U(1)_{B-L}$ originate from CHARM~\cite{CHARM:1985anb}, Orsay~\cite{Davier:1989wz}, and NuCal~\cite{Blumlein:1991xh} and were recasted via the \texttt{DarkCast} package~\cite{Ilten:2018crw}. Additonal bounds come from CHARM-II~\cite{CHARM-II:1993phx, Bilmis:2015lja}, NA64~\cite{NA64:2022yly} and FASER~\cite{FASER:2023tle}. The bounds for the $U(1)_{B}$ gauge boson originate from  NuCal~\cite{Blumlein:1991xh, Blumlein:1990ay}, KLOE~II~\cite{Anastasi:2015qla, KLOE-2:2012lii}, LHCb~\cite{LHCb:2017trq, LHCb:2019vmc}, and LSND~\cite{Bauer:2018onh} as summarized in Ref.~\cite{Foguel:2022ppx}. Since the model is anomalous, additional constraints coming from otherwise enhanced rare $K$, $B$ and $Z$ decays are shown in light gray following Ref.~\cite{Dror:2017ehi}. The recasted sensitivity of FASER's ALP search obtained using our form factors and the quasi-real approximation is shown as blue solid curve. For the $U(1)_{B}$ gauge boson, we also show the results using the the FWW approximation as presented by the FASER collaboration. While we see mild differences between the predictions, especially close to the $\omega$ resonance, the overall sensitivity stays roughly the same. 

\subsection{Protophobic Gauge Boson}

\begin{figure*}[t]
\centering
\includegraphics[width=0.49\textwidth]{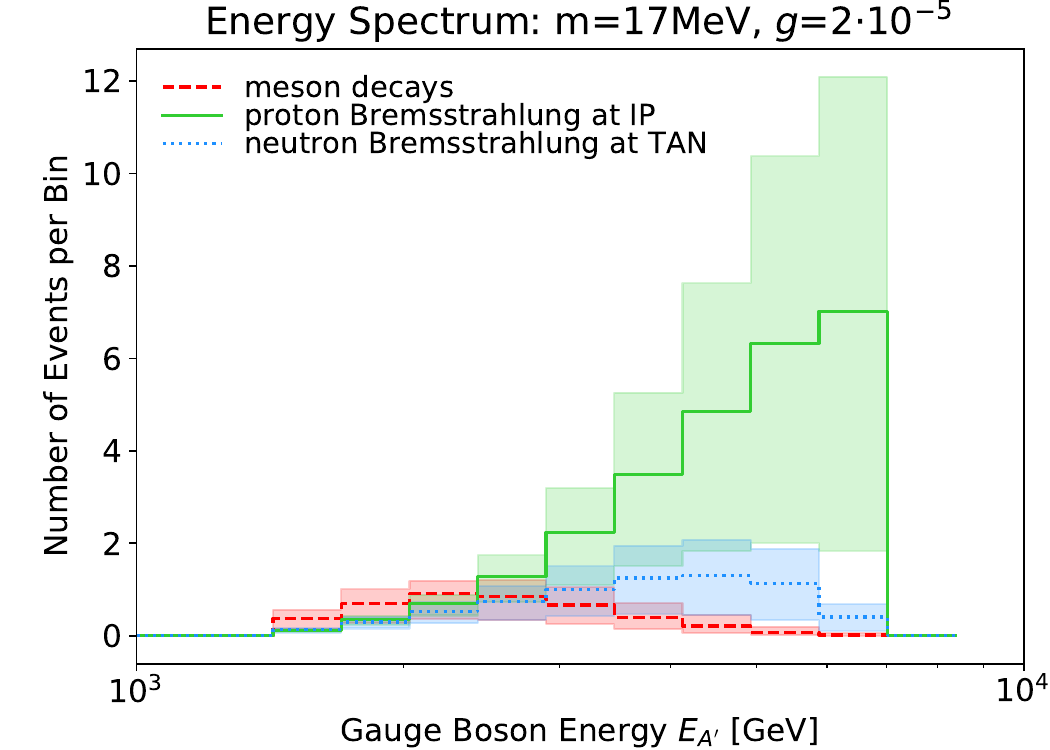} 
\includegraphics[width=0.49\textwidth]{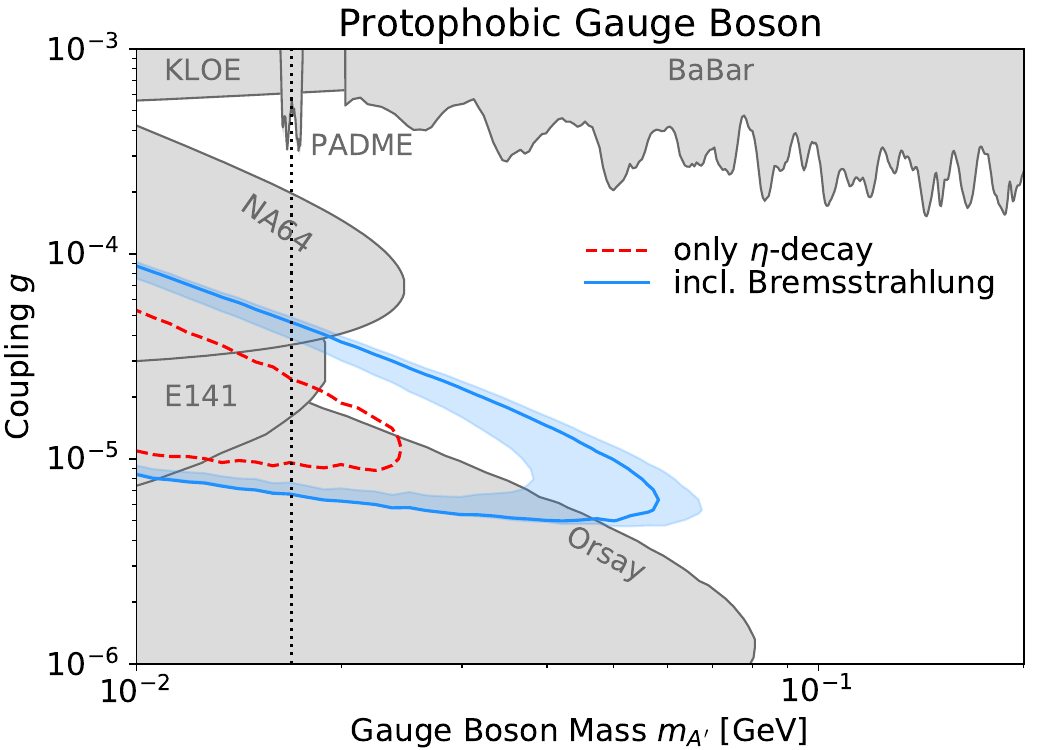} 
\caption{\textbf{Left:} Expected energy spectrum of a protophobic gauge boson with $m_{A'}=17$~MeV and $g=2\cdot 10^5$ at FASER. Contributions from meson decays, proton Bremsstrahlung at the IP and neutron Bremsstrahlung at the TAN are shown in red, green and blue, respectively. The shaded regions show the uncertainty band.  \textbf{Right:} Recasted sensitivity of FASER's ALP search in the protophobic gauge boson parameter space. The red dashed curve shows the bounds accounting only for their production via meson decays. The blue curve shows our recasted result including meson decays as well as proton Bremsstrahlung at the IP and neutron Bremsstrahlung at the TAN modeled via the QRA-D splitting functions and the new form factors. The gray shaded region corresponds to previously bounds.}
\label{fig:Protophobic}
\end{figure*}

A less explored yet intriguing scenario is the protophobic gauge boson~\cite{Feng:2016ysn, Feng:2020mbt}. This particle was originally proposed as an explanation of the $>6\sigma$ excess of electron–positron pairs at $17$ MeV reported in rare nuclear decays by the ATOMKI collaboration~\cite{Krasznahorkay:2015iga, Firak:2020eil}, and has since become the focus of numerous dedicated searches~\cite{Alves:2023ree}. In this model, the quark charges are orthogonal to those of the dark photon. As a consequence, the decay $\pi^0 \to A'\gamma$ is strongly suppressed~\cite{Ilten:2018crw} such that $\eta \to A'\gamma$ becomes the dominant production channel from meson decay.

One may naively assume that a protophobic gauge boson would not be produced in proton Bremsstrahlung. However, similar to the case of dark photon production via neutron Bremsstrahlung, this is not the case due to the finiteness of $F_1(t)$ for $t = m_{A'}^2 > 0$ and $F_2(t)$ for all $t$. In addition, the protophobic gauge boson can be effectively produced via neutron Bremsstrahlung as pointed out previously in Ref.~\cite{Dev:2023zts}. We show the production rate of a protophobic gauge boson in FASER's acceptance region as a function of mass as a red curve in the right panel of \cref{fig:prod_vs_mass}. We can see that the the production via proton Bremsstrahlung provides the dominant contribution. Protophobic gauge boson production via neutron Bremsstrahlung is suppressed in comparison, although less than in the case of dark photons. 

The left panel of \cref{fig:Protophobic} shows the energy spectrum of protophobic gauge bosons decaying in FASER for $m_{A'} = 17~\mev$ and $\epsilon = 2 \times 10^{-5}$. The contributions from $\eta$-meson decays, proton Bremsstrahlung at the primary interaction point, and neutron Bremsstrahlung at the TAN are shown as red dotted, green solid, and blue dotted lines, respectively. As before, the uncertainties in the Bremsstrahlung contributions are estimated by varying the form factor models, the scale $\Lambda_p$, and the neutron production model. We find that proton Bremsstrahlung provides the dominant production mode, followed by neutron Bremsstrahlung and meson decays.

The right panel of \cref{fig:Protophobic} shows the parameter space of the protophobic gauge boson. The gray shaded regions were previously excluded by searches at BaBar~\cite{BaBar:2014zli}, KLOE~\cite{KLOE-2:2018kqf},  PADME~\cite{Bossi:2025ptv},  E141~\cite{Riordan:1987aw},  NA64~\cite{NA64:2019auh}, and Orsay~\cite{Davier:1989wz}, as obtained via \texttt{DarkCast}. The colored lines show the constraints from recasting FASER's ALP search. The red dashed line assumed that the protophobic gauge boson is only produced in $\eta$-decay. The blue solid curve additionally includes proton and neutron Bremsstrahlung. We see that the inclusion of both Bremsstrahlung production modes significantly increases FASER's sensitivity in this model. The uncertainties on the reach, mainly associated with variations of the value of $\Lambda_p$, are found to be sizable. 

The vertical dotted line corresponds to a mass of 17~MeV, as needed to explain the signal observed by ATOMKI and more recently PADME~\cite{Arias-Aragon:2025wdt}. We can see that FASER's ALP search excludes the range of gauge coupling between $7\cdot 10^{-6}$ and $7 \cdot 10^{-5}$, hence validating the decade-old result from E141.  

\subsection{Millicharged Particles}

A millicharged particle $\chi$ is a new fermion carrying a fractional electric charge $Q_\chi = \epsilon e$ with $\epsilon < 1$. Such particles can arise as a low-energy manifestation of a coupling to a kinetically mixed massless dark photon~\cite{Holdom:1985ag}, or more simply as particles with a small hypercharge. Numerous accelerator and collider experiments have been proposed or are currently operating to search for millicharged particles.

Among them, the experiment with the strongest projected sensitivity is FORMOSA, which would be located in the far-forward region of the LHC~\cite{Foroughi-Abari:2020qar}. FORMOSA follows the principle design of the milliQan detector~\cite{Haas:2014dda, Ball:2016zrp, milliQan:2021lne} and consists of an array of plastic scintillators designed to identify millicharged particles through coincident signals of small energy deposits left along their trajectories. Previous studies of millicharged particle fluxes at FORMOSA have considered their production via pseudoscalar and vector meson decays as well as the Drell-Yan process a high masses. However, for GeV-scale masses, Bremsstrahlung production in $pp \to \gamma^* + X$ with subsequent $\gamma^* \to \chi \bar{\chi}$ is also expected to provide a relevant contribution, though it has so far been neglected.

\begin{figure*}[t]
\centering
\includegraphics[width=0.49\textwidth]{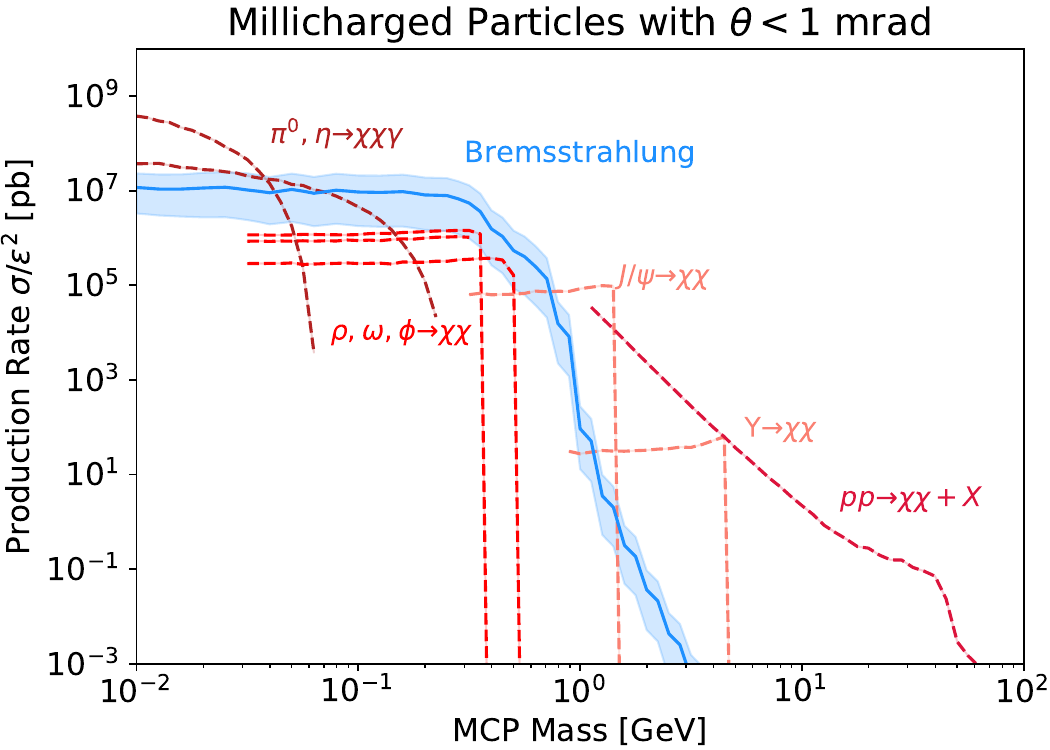} 
\includegraphics[width=0.49\textwidth]{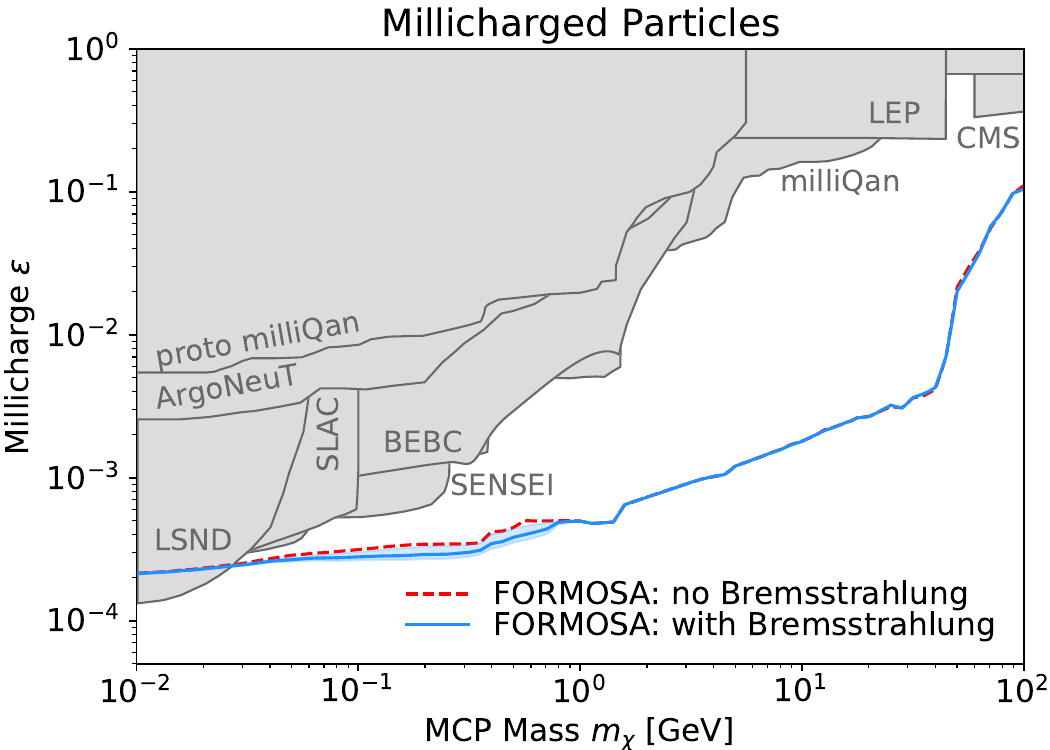} 
\caption{\textbf{Left:} Production cross section of
millicharged particles in the forward direction, with polar angles $\theta<1$~mrad, for different production modes and millicharged particle masses. Meson decays are shown as red dashed lines, while Bremsstrahlung is shown as blue solid line. The shaded region shows the corresponding uncertainty band. \textbf{Right:} Projected sensitivity of FORMOSA in the parameter space of millicharged particles. The red dashed curve corresponds to the sensitivity obtained when only considering meson decays and Drell-Yan production. The blue solid curve shows the projected sensitivity after accounting for millicharged particle production via proton Bremsstrahlung. The gray shaded region corresponds to previously bounds. }
\label{fig:MCP}
\end{figure*}

Following Ref.~\cite{Gninenko:2018ter} we can write the corresponding production rate as 
\be 
d\sigma_{pp \to \chi\chi + X}  &= \frac{\epsilon^2 e^2}{6 \pi^2} \int \frac{d m_{\gamma^*}}{m_{\gamma^*}} d\sigma_{pp \to \gamma^* + X}(m_{\gamma^*}) \\ 
& \ \  \times \ \Big(1-\frac{4 m_\chi^2}{m_{\gamma^*}^2}\Big)^{1/2} \Big(1+\frac{2 m_\chi^2}{m_{\gamma^*}^2}\Big) \ .
\ee
Since $\sigma_{pp \to \gamma^* + X}$ is equivalent to $\sigma_{pp \to A' + X}$ for $m_{A'} = m_{\gamma^*}$, we can reuse the dark photon production results with the QRA+D splitting function and the updated form factors. In practice, we perform a numerical integration over $m_{\gamma^*}$ and subsequently decay the virtual photon into $\chi\bar\chi$ to obtain the $\chi$ spectra.

The production rate of millicharged particles within 1~mrad around the beam collision axis, roughly corresponding to FORMOSA's coverage, is shown in the left panel of \cref{fig:MCP}. We can see that the Bremsstrahlung mode indeed dominates in the mass range  $100~\mev < m_\chi < 1~\gev$. The corresponding uncertainty band was obtained by varying the form factor model and the value of $\Lambda_p$ as before. 

The right panel of  \cref{fig:MCP} shows the parameter space of millicharged particles, spanned by their mass $m_\chi$ and the charge $\epsilon = Q_\chi / e$ as estimated in FORESEE. The gray shaded regions are already excluded by previously millicharged particle searches at BEBC~\cite{Marocco:2020dqu}, SLAC~\cite{Prinz:1998ua}, LEP~\cite{Davidson:2000hf,OPAL:1995uwx}, CMS~\cite{CMS:2012xi, Jaeckel:2012yz, CMS:2024eyx}, LSND~\cite{Magill:2018tbb}, ArgoNeuT~\cite{ArgoNeuT:2019ckq}, milliQan~\cite{Ball:2020dnx, Alcott:2025rxn}, and SENSEI~\cite{SENSEI:2023gie}. The red dashed line shows the reach without Bremsstrahlung, while the blue solid line shows the reach including Bremsstrahlung. We can see that the sensitivity in the sub-GeV region mildly improves.  

\section{Conclusion}
\label{sec:conclusion}

In this paper we have expanded the framework for probing GeV-scale dark vectors at colliders and fixed target experiments. The production model, based on initial state radiation, was generalized to account for both protons and neutrons, and a simple physically motivated model for the monopole and dipole form factors was proposed that provides a good fit to data and can be flexibly applied to a variety of dark vectors with non-universal couplings. We also introduced a modified $\chi^2$ function to better account for uncertainties in the fits to data. Adopting these form-factors in the QRA framework developed in Refs.~\cite{Foroughi-Abari:2021zbm, Foroughi-Abari:2024xlj}, exclusion contours were updated using the most recent data from FASER for the production and detection of dark photons and also $U(1)_{B-L}$, $U(1)_{B}$ and protophobic gauge bosons. Forecasts for millicharged particle sensitivity at FORMOSA at the FPF were also updated. The inclusion of this generic Bremsstrahlung production mode enhances sensitivity in the GeV mass range in all cases, Notably for FASER, the generalized model allowed for the first time an incorporation of neutron bremsstrahlung due to neutrons from the primary IP interacting downstream in the TAN.

The exploration of light dark sectors at colliders and fixed target experiments has matured and is entering a precision era with dedicated searches at a variety of proton beam experiments~\cite{Ilten:2022lfq} including FASER at the LHC, the FNAL short baseline program, and future experiments such as SHiP~\cite{Alekhin:2015byh, Ahdida:2023okr} and those at the planned FPF~\cite{Anchordoqui:2021ghd, Feng:2022inv, Adhikary:2024nlv}. The full exploitation of these facilities motivates improved modeling of production and detection and a quantitative assessment of uncertainties. The analysis in this paper is a further step in this direction, and this approach could be extended to other dark sector degrees of freedom such as Higgs portal scalars and ALPs. We also caution that there are in principle relevant production channels beyond initial state radiation. In particular, dark sector states can also be produced in final state radiation that in fully inelastic high-energy collisions will require a statistical treatment accounting for hadronization.

\vspace{4mm}
\section*{Acknowledgements}

We would like to thank Saeid Foroughi-Abari and Ekaterina Kriukova for useful discussions. The work of FK is supported in part by Heising-Simons Foundation Grant 2020-1840, by the U.S. National Science Foundation Grant PHY-2210283, and the Deutsche Forschungsgemeinschaft under Germany's Excellence Strategy -- EXC 2121 Quantum Universe -- 390833306. PR acknowledges financial support from the Fundação de Amparo à Pesquisa do Estado de São Paulo (FAPESP) under the contract 2020/10004-7. The work of PR and AR was also supported in part by NSERC, Canada.  FK further acknowledges support from the FAPESP Visiting Researcher Award Program (24/20695-8), which initiated this project.


\bibliography{literature}

\providecommand{\href}[2]{#2}\begingroup\raggedright\begin{thebibliography}{100}

\bibitem{Feng:2010gw}
J.~L. Feng, ``{Dark Matter Candidates from Particle Physics and Methods of Detection},'' \href{http://dx.doi.org/10.1146/annurev-astro-082708-101659}{{\em Ann. Rev. Astron. Astrophys.} {\bf 48} (2010)  495--545}, \href{http://arxiv.org/abs/1003.0904}{{\tt arXiv:1003.0904 [astro-ph.CO]}}.

\bibitem{Alexander:2016aln}
J.~Alexander {\em et al.}, ``{Dark Sectors 2016 Workshop: Community Report},'' \href{http://arxiv.org/abs/1608.08632}{{\tt arXiv:1608.08632 [hep-ph]}}.

\bibitem{Bertone:2018krk}
G.~Bertone and T.~Tait, M.~P., ``{A new era in the search for dark matter},'' \href{http://dx.doi.org/10.1038/s41586-018-0542-z}{{\em Nature} {\bf 562} (2018) no.~7725, 51--56}, \href{http://arxiv.org/abs/1810.01668}{{\tt arXiv:1810.01668 [astro-ph.CO]}}.

\bibitem{Arcadi:2017kky}
G.~Arcadi, M.~Dutra, P.~Ghosh, M.~Lindner, Y.~Mambrini, M.~Pierre, S.~Profumo, and F.~S. Queiroz, ``{The waning of the WIMP? A review of models, searches, and constraints},'' \href{http://dx.doi.org/10.1140/epjc/s10052-018-5662-y}{{\em Eur. Phys. J. C} {\bf 78} (2018) no.~3, 203}, \href{http://arxiv.org/abs/1703.07364}{{\tt arXiv:1703.07364 [hep-ph]}}.

\bibitem{Roszkowski:2017nbc}
L.~Roszkowski, E.~M. Sessolo, and S.~Trojanowski, ``{WIMP dark matter candidates and searches{\textemdash}current status and future prospects},'' \href{http://dx.doi.org/10.1088/1361-6633/aab913}{{\em Rept. Prog. Phys.} {\bf 81} (2018) no.~6, 066201}, \href{http://arxiv.org/abs/1707.06277}{{\tt arXiv:1707.06277 [hep-ph]}}.

\bibitem{Pospelov:2008zw}
M.~Pospelov, ``{Secluded U(1) below the weak scale},'' \href{http://dx.doi.org/10.1103/PhysRevD.80.095002}{{\em Phys. Rev. D} {\bf 80} (2009)  095002}, \href{http://arxiv.org/abs/0811.1030}{{\tt arXiv:0811.1030 [hep-ph]}}.

\bibitem{Batell:2009yf}
B.~Batell, M.~Pospelov, and A.~Ritz, ``{Probing a Secluded U(1) at B-factories},'' \href{http://dx.doi.org/10.1103/PhysRevD.79.115008}{{\em Phys. Rev. D} {\bf 79} (2009)  115008}, \href{http://arxiv.org/abs/0903.0363}{{\tt arXiv:0903.0363 [hep-ph]}}.

\bibitem{Essig:2009nc}
R.~Essig, P.~Schuster, and N.~Toro, ``{Probing Dark Forces and Light Hidden Sectors at Low-Energy e+e- Colliders},'' \href{http://dx.doi.org/10.1103/PhysRevD.80.015003}{{\em Phys. Rev. D} {\bf 80} (2009)  015003}, \href{http://arxiv.org/abs/0903.3941}{{\tt arXiv:0903.3941 [hep-ph]}}.

\bibitem{Reece:2009un}
M.~Reece and L.-T. Wang, ``{Searching for the light dark gauge boson in GeV-scale experiments},'' \href{http://dx.doi.org/10.1088/1126-6708/2009/07/051}{{\em JHEP} {\bf 07} (2009)  051}, \href{http://arxiv.org/abs/0904.1743}{{\tt arXiv:0904.1743 [hep-ph]}}.

\bibitem{Freytsis:2009bh}
M.~Freytsis, G.~Ovanesyan, and J.~Thaler, ``{Dark Force Detection in Low Energy e-p Collisions},'' \href{http://dx.doi.org/10.1007/JHEP01(2010)111}{{\em JHEP} {\bf 01} (2010)  111}, \href{http://arxiv.org/abs/0909.2862}{{\tt arXiv:0909.2862 [hep-ph]}}.

\bibitem{Batell:2009jf}
B.~Batell, M.~Pospelov, and A.~Ritz, ``{Multi-lepton Signatures of a Hidden Sector in Rare B Decays},'' \href{http://dx.doi.org/10.1103/PhysRevD.83.054005}{{\em Phys. Rev. D} {\bf 83} (2011)  054005}, \href{http://arxiv.org/abs/0911.4938}{{\tt arXiv:0911.4938 [hep-ph]}}.

\bibitem{Freytsis:2009ct}
M.~Freytsis, Z.~Ligeti, and J.~Thaler, ``{Constraining the Axion Portal with $B \to K l^+ l^-$},'' \href{http://dx.doi.org/10.1103/PhysRevD.81.034001}{{\em Phys. Rev. D} {\bf 81} (2010)  034001}, \href{http://arxiv.org/abs/0911.5355}{{\tt arXiv:0911.5355 [hep-ph]}}.

\bibitem{Essig:2010xa}
R.~Essig, P.~Schuster, N.~Toro, and B.~Wojtsekhowski, ``{An Electron Fixed Target Experiment to Search for a New Vector Boson A' Decaying to e+e-},'' \href{http://dx.doi.org/10.1007/JHEP02(2011)009}{{\em JHEP} {\bf 02} (2011)  009}, \href{http://arxiv.org/abs/1001.2557}{{\tt arXiv:1001.2557 [hep-ph]}}.

\bibitem{Essig:2010gu}
R.~Essig, R.~Harnik, J.~Kaplan, and N.~Toro, ``{Discovering New Light States at Neutrino Experiments},'' \href{http://dx.doi.org/10.1103/PhysRevD.82.113008}{{\em Phys. Rev. D} {\bf 82} (2010)  113008}, \href{http://arxiv.org/abs/1008.0636}{{\tt arXiv:1008.0636 [hep-ph]}}.

\bibitem{McDonald:2010fe}
K.~L. McDonald and D.~E. Morrissey, ``{Low-Energy Signals from Kinetic Mixing with a Warped Abelian Hidden Sector},'' \href{http://dx.doi.org/10.1007/JHEP02(2011)087}{{\em JHEP} {\bf 02} (2011)  087}, \href{http://arxiv.org/abs/1010.5999}{{\tt arXiv:1010.5999 [hep-ph]}}.

\bibitem{Williams:2011qb}
M.~Williams, C.~P. Burgess, A.~Maharana, and F.~Quevedo, ``{New Constraints (and Motivations) for Abelian Gauge Bosons in the MeV-TeV Mass Range},'' \href{http://dx.doi.org/10.1007/JHEP08(2011)106}{{\em JHEP} {\bf 08} (2011)  106}, \href{http://arxiv.org/abs/1103.4556}{{\tt arXiv:1103.4556 [hep-ph]}}.

\bibitem{Davoudiasl:2012ag}
H.~Davoudiasl, H.-S. Lee, and W.~J. Marciano, ``{'Dark' Z implications for Parity Violation, Rare Meson Decays, and Higgs Physics},'' \href{http://dx.doi.org/10.1103/PhysRevD.85.115019}{{\em Phys. Rev. D} {\bf 85} (2012)  115019}, \href{http://arxiv.org/abs/1203.2947}{{\tt arXiv:1203.2947 [hep-ph]}}.

\bibitem{Kahn:2012br}
Y.~Kahn and J.~Thaler, ``{Searching for an invisible A' vector boson with DarkLight},'' \href{http://dx.doi.org/10.1103/PhysRevD.86.115012}{{\em Phys. Rev. D} {\bf 86} (2012)  115012}, \href{http://arxiv.org/abs/1209.0777}{{\tt arXiv:1209.0777 [hep-ph]}}.

\bibitem{Andreas:2012mt}
S.~Andreas, C.~Niebuhr, and A.~Ringwald, ``{New Limits on Hidden Photons from Past Electron Beam Dumps},'' \href{http://dx.doi.org/10.1103/PhysRevD.86.095019}{{\em Phys. Rev. D} {\bf 86} (2012)  095019}, \href{http://arxiv.org/abs/1209.6083}{{\tt arXiv:1209.6083 [hep-ph]}}.

\bibitem{Battaglieri:2017aum}
M.~Battaglieri {\em et al.}, ``{US Cosmic Visions: New Ideas in Dark Matter 2017: Community Report},'' in {\em {U.S. Cosmic Visions: New Ideas in Dark Matter}}.
\newblock 7, 2017.
\newblock \href{http://arxiv.org/abs/1707.04591}{{\tt arXiv:1707.04591 [hep-ph]}}.

\bibitem{Knapen:2017xzo}
S.~Knapen, T.~Lin, and K.~M. Zurek, ``{Light Dark Matter: Models and Constraints},'' \href{http://dx.doi.org/10.1103/PhysRevD.96.115021}{{\em Phys. Rev. D} {\bf 96} (2017) no.~11, 115021}, \href{http://arxiv.org/abs/1709.07882}{{\tt arXiv:1709.07882 [hep-ph]}}.

\bibitem{Batell:2009di}
B.~Batell, M.~Pospelov, and A.~Ritz, ``{Exploring Portals to a Hidden Sector Through Fixed Targets},'' \href{http://dx.doi.org/10.1103/PhysRevD.80.095024}{{\em Phys. Rev. D} {\bf 80} (2009)  095024}, \href{http://arxiv.org/abs/0906.5614}{{\tt arXiv:0906.5614 [hep-ph]}}.

\bibitem{deNiverville:2011it}
P.~deNiverville, M.~Pospelov, and A.~Ritz, ``{Observing a light dark matter beam with neutrino experiments},'' \href{http://dx.doi.org/10.1103/PhysRevD.84.075020}{{\em Phys. Rev. D} {\bf 84} (2011)  075020}, \href{http://arxiv.org/abs/1107.4580}{{\tt arXiv:1107.4580 [hep-ph]}}.

\bibitem{deNiverville:2012ij}
P.~deNiverville, D.~McKeen, and A.~Ritz, ``{Signatures of sub-GeV dark matter beams at neutrino experiments},'' \href{http://dx.doi.org/10.1103/PhysRevD.86.035022}{{\em Phys. Rev. D} {\bf 86} (2012)  035022}, \href{http://arxiv.org/abs/1205.3499}{{\tt arXiv:1205.3499 [hep-ph]}}.

\bibitem{Kahn:2014sra}
Y.~Kahn, G.~Krnjaic, J.~Thaler, and M.~Toups, ``{DAE{\ensuremath{\delta}}ALUS and dark matter detection},'' \href{http://dx.doi.org/10.1103/PhysRevD.91.055006}{{\em Phys. Rev. D} {\bf 91} (2015) no.~5, 055006}, \href{http://arxiv.org/abs/1411.1055}{{\tt arXiv:1411.1055 [hep-ph]}}.

\bibitem{Adams:2013qkq}
{\bf LBNE} Collaboration, C.~Adams {\em et al.}, ``{The Long-Baseline Neutrino Experiment: Exploring Fundamental Symmetries of the Universe},'' in {\em {Snowmass 2013}: {Workshop on Energy Frontier}}.
\newblock 7, 2013.
\newblock \href{http://arxiv.org/abs/1307.7335}{{\tt arXiv:1307.7335 [hep-ex]}}.

\bibitem{Soper:2014ska}
D.~E. Soper, M.~Spannowsky, C.~J. Wallace, and T.~M.~P. Tait, ``{Scattering of Dark Particles with Light Mediators},'' \href{http://dx.doi.org/10.1103/PhysRevD.90.115005}{{\em Phys. Rev. D} {\bf 90} (2014) no.~11, 115005}, \href{http://arxiv.org/abs/1407.2623}{{\tt arXiv:1407.2623 [hep-ph]}}.

\bibitem{Dobrescu:2014ita}
B.~A. Dobrescu and C.~Frugiuele, ``{GeV-Scale Dark Matter: Production at the Main Injector},'' \href{http://dx.doi.org/10.1007/JHEP02(2015)019}{{\em JHEP} {\bf 02} (2015)  019}, \href{http://arxiv.org/abs/1410.1566}{{\tt arXiv:1410.1566 [hep-ph]}}.

\bibitem{Coloma:2015pih}
P.~Coloma, B.~A. Dobrescu, C.~Frugiuele, and R.~Harnik, ``{Dark matter beams at LBNF},'' \href{http://dx.doi.org/10.1007/JHEP04(2016)047}{{\em JHEP} {\bf 04} (2016)  047}, \href{http://arxiv.org/abs/1512.03852}{{\tt arXiv:1512.03852 [hep-ph]}}.

\bibitem{Alpigiani:2018fgd}
{\bf MATHUSLA} Collaboration, C.~Alpigiani {\em et al.}, ``{A Letter of Intent for MATHUSLA: A Dedicated Displaced Vertex Detector above ATLAS or CMS.},'' \href{http://arxiv.org/abs/1811.00927}{{\tt arXiv:1811.00927 [physics.ins-det]}}.

\bibitem{Ariga:2018pin}
{\bf FASER} Collaboration, A.~Ariga {\em et al.}, ``{Technical Proposal for FASER: ForwArd Search ExpeRiment at the LHC},'' \href{http://arxiv.org/abs/1812.09139}{{\tt arXiv:1812.09139 [physics.ins-det]}}.

\bibitem{Dutta:2020vop}
B.~Dutta, D.~Kim, S.~Liao, J.-C. Park, S.~Shin, L.~E. Strigari, and A.~Thompson, ``{Searching for dark matter signals in timing spectra at neutrino experiments},'' \href{http://dx.doi.org/10.1007/JHEP01(2022)144}{{\em JHEP} {\bf 01} (2022)  144}, \href{http://arxiv.org/abs/2006.09386}{{\tt arXiv:2006.09386 [hep-ph]}}.

\bibitem{Batell:2021blf}
B.~Batell, J.~L. Feng, and S.~Trojanowski, ``{Detecting Dark Matter with Far-Forward Emulsion and Liquid Argon Detectors at the LHC},'' \href{http://dx.doi.org/10.1103/PhysRevD.103.075023}{{\em Phys. Rev. D} {\bf 103} (2021) no.~7, 075023}, \href{http://arxiv.org/abs/2101.10338}{{\tt arXiv:2101.10338 [hep-ph]}}.

\bibitem{Batell:2021aja}
B.~Batell, J.~L. Feng, A.~Ismail, F.~Kling, R.~M. Abraham, and S.~Trojanowski, ``{Discovering dark matter at the LHC through its nuclear scattering in far-forward emulsion and liquid argon detectors},'' \href{http://dx.doi.org/10.1103/PhysRevD.104.035036}{{\em Phys. Rev. D} {\bf 104} (2021) no.~3, 035036}, \href{http://arxiv.org/abs/2107.00666}{{\tt arXiv:2107.00666 [hep-ph]}}.

\bibitem{Bjorken:2009mm}
J.~D. Bjorken, R.~Essig, P.~Schuster, and N.~Toro, ``{New Fixed-Target Experiments to Search for Dark Gauge Forces},'' \href{http://dx.doi.org/10.1103/PhysRevD.80.075018}{{\em Phys. Rev. D} {\bf 80} (2009)  075018}, \href{http://arxiv.org/abs/0906.0580}{{\tt arXiv:0906.0580 [hep-ph]}}.

\bibitem{Izaguirre:2013uxa}
E.~Izaguirre, G.~Krnjaic, P.~Schuster, and N.~Toro, ``{New Electron Beam-Dump Experiments to Search for MeV to few-GeV Dark Matter},'' \href{http://dx.doi.org/10.1103/PhysRevD.88.114015}{{\em Phys. Rev. D} {\bf 88} (2013)  114015}, \href{http://arxiv.org/abs/1307.6554}{{\tt arXiv:1307.6554 [hep-ph]}}.

\bibitem{Diamond:2013oda}
M.~D. Diamond and P.~Schuster, ``{Searching for Light Dark Matter with the SLAC Millicharge Experiment},'' \href{http://dx.doi.org/10.1103/PhysRevLett.111.221803}{{\em Phys. Rev. Lett.} {\bf 111} (2013) no.~22, 221803}, \href{http://arxiv.org/abs/1307.6861}{{\tt arXiv:1307.6861 [hep-ph]}}.

\bibitem{Izaguirre:2014dua}
E.~Izaguirre, G.~Krnjaic, P.~Schuster, and N.~Toro, ``{Physics motivation for a pilot dark matter search at Jefferson Laboratory},'' \href{http://dx.doi.org/10.1103/PhysRevD.90.014052}{{\em Phys. Rev. D} {\bf 90} (2014) no.~1, 014052}, \href{http://arxiv.org/abs/1403.6826}{{\tt arXiv:1403.6826 [hep-ph]}}.

\bibitem{Batell:2014mga}
B.~Batell, R.~Essig, and Z.~Surujon, ``{Strong Constraints on Sub-GeV Dark Sectors from SLAC Beam Dump E137},'' \href{http://dx.doi.org/10.1103/PhysRevLett.113.171802}{{\em Phys. Rev. Lett.} {\bf 113} (2014) no.~17, 171802}, \href{http://arxiv.org/abs/1406.2698}{{\tt arXiv:1406.2698 [hep-ph]}}.

\bibitem{Lees:2017lec}
{\bf BaBar} Collaboration, J.~P. Lees {\em et al.}, ``{Search for Invisible Decays of a Dark Photon Produced in ${e}^{+}{e}^{-}$ Collisions at BaBar},'' \href{http://dx.doi.org/10.1103/PhysRevLett.119.131804}{{\em Phys. Rev. Lett.} {\bf 119} (2017) no.~13, 131804}, \href{http://arxiv.org/abs/1702.03327}{{\tt arXiv:1702.03327 [hep-ex]}}.

\bibitem{Berlin:2020uwy}
A.~Berlin, P.~deNiverville, A.~Ritz, P.~Schuster, and N.~Toro, ``{Sub-GeV dark matter production at fixed-target experiments},'' \href{http://dx.doi.org/10.1103/PhysRevD.102.095011}{{\em Phys. Rev. D} {\bf 102} (2020) no.~9, 095011}, \href{http://arxiv.org/abs/2003.03379}{{\tt arXiv:2003.03379 [hep-ph]}}.

\bibitem{Krnjaic:2022ozp}
G.~Krnjaic {\em et al.}, ``{A Snowmass Whitepaper: Dark Matter Production at Intensity-Frontier Experiments},'' \href{http://arxiv.org/abs/2207.00597}{{\tt arXiv:2207.00597 [hep-ph]}}.

\bibitem{Berlin:2018bsc}
A.~Berlin, N.~Blinov, G.~Krnjaic, P.~Schuster, and N.~Toro, ``{Dark Matter, Millicharges, Axion and Scalar Particles, Gauge Bosons, and Other New Physics with LDMX},'' \href{http://dx.doi.org/10.1103/PhysRevD.99.075001}{{\em Phys. Rev. D} {\bf 99} (2019) no.~7, 075001}, \href{http://arxiv.org/abs/1807.01730}{{\tt arXiv:1807.01730 [hep-ph]}}.

\bibitem{Bauer:2018onh}
M.~Bauer, P.~Foldenauer, and J.~Jaeckel, ``{Hunting All the Hidden Photons},'' \href{http://dx.doi.org/10.1007/JHEP07(2018)094}{{\em JHEP} {\bf 07} (2018)  094}, \href{http://arxiv.org/abs/1803.05466}{{\tt arXiv:1803.05466 [hep-ph]}}.

\bibitem{Berlin:2023qco}
A.~Berlin, G.~Krnjaic, and E.~Pinetti, ``{Reviving MeV-GeV indirect detection with inelastic dark matter},'' \href{http://dx.doi.org/10.1103/PhysRevD.110.035015}{{\em Phys. Rev. D} {\bf 110} (2024) no.~3, 035015}, \href{http://arxiv.org/abs/2311.00032}{{\tt arXiv:2311.00032 [hep-ph]}}.

\bibitem{Lu:2023cet}
C.-T. Lu, J.~Tu, and L.~Wu, ``{Probing inelastic dark matter at the LHC, FASER, and STCF},'' \href{http://dx.doi.org/10.1103/PhysRevD.109.015018}{{\em Phys. Rev. D} {\bf 109} (2024) no.~1, 015018}, \href{http://arxiv.org/abs/2309.00271}{{\tt arXiv:2309.00271 [hep-ph]}}.

\bibitem{Mongillo:2023hbs}
M.~Mongillo, A.~Abdullahi, B.~B. Oberhauser, P.~Crivelli, M.~Hostert, D.~Massaro, L.~M. Bueno, and S.~Pascoli, ``{Constraining light thermal inelastic dark matter with NA64},'' \href{http://dx.doi.org/10.1140/epjc/s10052-023-11536-5}{{\em Eur. Phys. J. C} {\bf 83} (2023) no.~5, 391}, \href{http://arxiv.org/abs/2302.05414}{{\tt arXiv:2302.05414 [hep-ph]}}.

\bibitem{Filimonova:2022pkj}
A.~Filimonova, S.~Junius, L.~Lopez~Honorez, and S.~Westhoff, ``{Inelastic Dirac dark matter},'' \href{http://dx.doi.org/10.1007/JHEP06(2022)048}{{\em JHEP} {\bf 06} (2022)  048}, \href{http://arxiv.org/abs/2201.08409}{{\tt arXiv:2201.08409 [hep-ph]}}.

\bibitem{CarrilloGonzalez:2021lxm}
M.~Carrillo~Gonz{\'a}lez and N.~Toro, ``{Cosmology and signals of light pseudo-Dirac dark matter},'' \href{http://dx.doi.org/10.1007/JHEP04(2022)060}{{\em JHEP} {\bf 04} (2022)  060}, \href{http://arxiv.org/abs/2108.13422}{{\tt arXiv:2108.13422 [hep-ph]}}.

\bibitem{Batell:2021snh}
B.~Batell, J.~L. Feng, M.~Fieg, A.~Ismail, F.~Kling, R.~M. Abraham, and S.~Trojanowski, ``{Hadrophilic dark sectors at the Forward Physics Facility},'' \href{http://dx.doi.org/10.1103/PhysRevD.105.075001}{{\em Phys. Rev. D} {\bf 105} (2022) no.~7, 075001}, \href{http://arxiv.org/abs/2111.10343}{{\tt arXiv:2111.10343 [hep-ph]}}.

\bibitem{Garcia:2024uwf}
G.~D.~V. Garcia, F.~Kahlhoefer, M.~Ovchynnikov, and T.~Schwetz, ``{Not-so-inelastic Dark Matter},'' \href{http://dx.doi.org/10.1007/JHEP02(2025)127}{{\em JHEP} {\bf 02} (2025)  127}, \href{http://arxiv.org/abs/2405.08081}{{\tt arXiv:2405.08081 [hep-ph]}}.

\bibitem{APEX:2011dww}
{\bf APEX} Collaboration, S.~Abrahamyan {\em et al.}, ``{Search for a New Gauge Boson in Electron-Nucleus Fixed-Target Scattering by the APEX Experiment},'' \href{http://dx.doi.org/10.1103/PhysRevLett.107.191804}{{\em Phys. Rev. Lett.} {\bf 107} (2011)  191804}, \href{http://arxiv.org/abs/1108.2750}{{\tt arXiv:1108.2750 [hep-ex]}}.

\bibitem{KLOE-2:2011hhj}
{\bf KLOE-2} Collaboration, F.~Archilli {\em et al.}, ``{Search for a vector gauge boson in $\phi$ meson decays with the KLOE detector},'' \href{http://dx.doi.org/10.1016/j.physletb.2011.11.033}{{\em Phys. Lett. B} {\bf 706} (2012)  251--255}, \href{http://arxiv.org/abs/1110.0411}{{\tt arXiv:1110.0411 [hep-ex]}}.

\bibitem{BaBar:2012bkw}
{\bf BaBar} Collaboration, J.~P. Lees {\em et al.}, ``{Search for Low-Mass Dark-Sector Higgs Bosons},'' \href{http://dx.doi.org/10.1103/PhysRevLett.108.211801}{{\em Phys. Rev. Lett.} {\bf 108} (2012)  211801}, \href{http://arxiv.org/abs/1202.1313}{{\tt arXiv:1202.1313 [hep-ex]}}.

\bibitem{Beacham:2019nyx}
J.~Beacham {\em et al.}, ``{Physics Beyond Colliders at CERN: Beyond the Standard Model Working Group Report},'' \href{http://dx.doi.org/10.1088/1361-6471/ab4cd2}{{\em J. Phys. G} {\bf 47} (2020) no.~1, 010501}, \href{http://arxiv.org/abs/1901.09966}{{\tt arXiv:1901.09966 [hep-ex]}}.

\bibitem{LBNE:2013dhi}
{\bf LBNE} Collaboration, C.~Adams {\em et al.}, ``{The Long-Baseline Neutrino Experiment: Exploring Fundamental Symmetries of the Universe},'' in {\em {Snowmass 2013}: {Workshop on Energy Frontier}}.
\newblock 7, 2013.
\newblock \href{http://arxiv.org/abs/1307.7335}{{\tt arXiv:1307.7335 [hep-ex]}}.

\bibitem{MiniBooNE:2017nqe}
{\bf MiniBooNE} Collaboration, A.~A. Aguilar-Arevalo {\em et al.}, ``{Dark Matter Search in a Proton Beam Dump with MiniBooNE},'' \href{http://dx.doi.org/10.1103/PhysRevLett.118.221803}{{\em Phys. Rev. Lett.} {\bf 118} (2017) no.~22, 221803}, \href{http://arxiv.org/abs/1702.02688}{{\tt arXiv:1702.02688 [hep-ex]}}.

\bibitem{MiniBooNEDM:2018cxm}
{\bf MiniBooNE DM} Collaboration, A.~A. Aguilar-Arevalo {\em et al.}, ``{Dark Matter Search in Nucleon, Pion, and Electron Channels from a Proton Beam Dump with MiniBooNE},'' \href{http://dx.doi.org/10.1103/PhysRevD.98.112004}{{\em Phys. Rev. D} {\bf 98} (2018) no.~11, 112004}, \href{http://arxiv.org/abs/1807.06137}{{\tt arXiv:1807.06137 [hep-ex]}}.

\bibitem{MATHUSLA:2018bqv}
{\bf MATHUSLA} Collaboration, C.~Alpigiani {\em et al.}, ``{A Letter of Intent for MATHUSLA: A Dedicated Displaced Vertex Detector above ATLAS or CMS.},'' \href{http://arxiv.org/abs/1811.00927}{{\tt arXiv:1811.00927 [physics.ins-det]}}.

\bibitem{BaBar:2017tiz}
{\bf BaBar} Collaboration, J.~P. Lees {\em et al.}, ``{Search for Invisible Decays of a Dark Photon Produced in ${e}^{+}{e}^{-}$ Collisions at BaBar},'' \href{http://dx.doi.org/10.1103/PhysRevLett.119.131804}{{\em Phys. Rev. Lett.} {\bf 119} (2017) no.~13, 131804}, \href{http://arxiv.org/abs/1702.03327}{{\tt arXiv:1702.03327 [hep-ex]}}.

\bibitem{NA64:2019imj}
D.~Banerjee {\em et al.}, ``{Dark matter search in missing energy events with NA64},'' \href{http://dx.doi.org/10.1103/PhysRevLett.123.121801}{{\em Phys. Rev. Lett.} {\bf 123} (2019) no.~12, 121801}, \href{http://arxiv.org/abs/1906.00176}{{\tt arXiv:1906.00176 [hep-ex]}}.

\bibitem{Antel:2023hkf}
C.~Antel {\em et al.}, ``{Feebly-interacting particles: FIPs 2022 Workshop Report},'' \href{http://dx.doi.org/10.1140/epjc/s10052-023-12168-5}{{\em Eur. Phys. J. C} {\bf 83} (2023) no.~12, 1122}, \href{http://arxiv.org/abs/2305.01715}{{\tt arXiv:2305.01715 [hep-ph]}}.

\bibitem{Berlin:2018jbm}
A.~Berlin and F.~Kling, ``{Inelastic Dark Matter at the LHC Lifetime Frontier: ATLAS, CMS, LHCb, CODEX-b, FASER, and MATHUSLA},'' \href{http://dx.doi.org/10.1103/PhysRevD.99.015021}{{\em Phys. Rev. D} {\bf 99} (2019) no.~1, 015021}, \href{http://arxiv.org/abs/1810.01879}{{\tt arXiv:1810.01879 [hep-ph]}}.

\bibitem{Feng:2017uoz}
J.~L. Feng, I.~Galon, F.~Kling, and S.~Trojanowski, ``{ForwArd Search ExpeRiment at the LHC},'' \href{http://dx.doi.org/10.1103/PhysRevD.97.035001}{{\em Phys. Rev. D} {\bf 97} (2018) no.~3, 035001}, \href{http://arxiv.org/abs/1708.09389}{{\tt arXiv:1708.09389 [hep-ph]}}.

\bibitem{SHiP:2015vad}
{\bf SHiP} Collaboration, M.~Anelli {\em et al.}, ``{A facility to Search for Hidden Particles (SHiP) at the CERN SPS},'' \href{http://arxiv.org/abs/1504.04956}{{\tt arXiv:1504.04956 [physics.ins-det]}}.

\bibitem{Machado:2019oxb}
P.~A. Machado, O.~Palamara, and D.~W. Schmitz, ``{The Short-Baseline Neutrino Program at Fermilab},'' \href{http://dx.doi.org/10.1146/annurev-nucl-101917-020949}{{\em Ann. Rev. Nucl. Part. Sci.} {\bf 69} (2019)  363--387}, \href{http://arxiv.org/abs/1903.04608}{{\tt arXiv:1903.04608 [hep-ex]}}.

\bibitem{DUNE:2020fgq}
{\bf DUNE} Collaboration, B.~Abi {\em et al.}, ``{Prospects for beyond the Standard Model physics searches at the Deep Underground Neutrino Experiment},'' \href{http://dx.doi.org/10.1140/epjc/s10052-021-09007-w}{{\em Eur. Phys. J. C} {\bf 81} (2021) no.~4, 322}, \href{http://arxiv.org/abs/2008.12769}{{\tt arXiv:2008.12769 [hep-ex]}}.

\bibitem{Celentano:2020vtu}
A.~Celentano, L.~Darm{\'e}, L.~Marsicano, and E.~Nardi, ``{New production channels for light dark matter in hadronic showers},'' \href{http://dx.doi.org/10.1103/PhysRevD.102.075026}{{\em Phys. Rev. D} {\bf 102} (2020) no.~7, 075026}, \href{http://arxiv.org/abs/2006.09419}{{\tt arXiv:2006.09419 [hep-ph]}}.

\bibitem{Capozzi:2021nmp}
F.~Capozzi, B.~Dutta, G.~Gurung, W.~Jang, I.~M. Shoemaker, A.~Thompson, and J.~Yu, ``{Extending the reach of leptophilic boson searches at DUNE and MiniBooNE with bremsstrahlung and resonant production},'' \href{http://dx.doi.org/10.1103/PhysRevD.104.115010}{{\em Phys. Rev. D} {\bf 104} (2021) no.~11, 115010}, \href{http://arxiv.org/abs/2108.03262}{{\tt arXiv:2108.03262 [hep-ph]}}.

\bibitem{Blinov:2024pza}
N.~Blinov, P.~J. Fox, K.~J. Kelly, P.~A.~N. Machado, and R.~Plestid, ``{Dark fluxes from electromagnetic cascades},'' \href{http://dx.doi.org/10.1007/JHEP07(2024)022}{{\em JHEP} {\bf 07} (2024)  022}, \href{http://arxiv.org/abs/2401.06843}{{\tt arXiv:2401.06843 [hep-ph]}}.

\bibitem{LoChiatto:2024guj}
P.~Lo~Chiatto and F.~Yu, ``{Consistent electroweak phenomenology of a nearly degenerate Z' boson},'' \href{http://dx.doi.org/10.1103/PhysRevD.111.035001}{{\em Phys. Rev. D} {\bf 111} (2025) no.~3, 035001}, \href{http://arxiv.org/abs/2405.03396}{{\tt arXiv:2405.03396 [hep-ph]}}.

\bibitem{Altmannshofer:2022ckw}
W.~Altmannshofer, J.~A. Dror, and S.~Gori, ``{New Opportunities for Detecting Axion-Lepton Interactions},'' \href{http://dx.doi.org/10.1103/PhysRevLett.130.241801}{{\em Phys. Rev. Lett.} {\bf 130} (2023) no.~24, 241801}, \href{http://arxiv.org/abs/2209.00665}{{\tt arXiv:2209.00665 [hep-ph]}}.

\bibitem{Curtin:2023bcf}
D.~Curtin, Y.~Kahn, and R.~Nguyen, ``{Dark photons from charged pion bremsstrahlung at proton beam experiments},'' \href{http://dx.doi.org/10.1103/PhysRevD.108.095039}{{\em Phys. Rev. D} {\bf 108} (2023) no.~9, 095039}, \href{http://arxiv.org/abs/2305.19309}{{\tt arXiv:2305.19309 [hep-ph]}}.

\bibitem{Kyselov:2024dmi}
Y.~Kyselov and M.~Ovchynnikov, ``{Searches for long-lived dark photons at proton accelerator experiments},'' \href{http://dx.doi.org/10.1103/PhysRevD.111.015030}{{\em Phys. Rev. D} {\bf 111} (2025) no.~1, 015030}, \href{http://arxiv.org/abs/2409.11096}{{\tt arXiv:2409.11096 [hep-ph]}}.

\bibitem{Anchordoqui:2021ghd}
L.~A. Anchordoqui {\em et al.}, ``{The Forward Physics Facility: Sites, experiments, and physics potential},'' \href{http://dx.doi.org/10.1016/j.physrep.2022.04.004}{{\em Phys. Rept.} {\bf 968} (2022)  1--50}, \href{http://arxiv.org/abs/2109.10905}{{\tt arXiv:2109.10905 [hep-ph]}}.

\bibitem{Feng:2022inv}
J.~L. Feng {\em et al.}, ``{The Forward Physics Facility at the High-Luminosity LHC},'' \href{http://dx.doi.org/10.1088/1361-6471/ac865e}{{\em J. Phys. G} {\bf 50} (2023) no.~3, 030501}, \href{http://arxiv.org/abs/2203.05090}{{\tt arXiv:2203.05090 [hep-ex]}}.

\bibitem{Adhikary:2024nlv}
J.~Adhikary {\em et al.}, ``{Scientific program for the Forward Physics Facility},'' \href{http://dx.doi.org/10.1140/epjc/s10052-025-14048-6}{{\em Eur. Phys. J. C} {\bf 85} (2025) no.~4, 430}, \href{http://arxiv.org/abs/2411.04175}{{\tt arXiv:2411.04175 [hep-ex]}}.

\bibitem{Blumlein:2013cua}
J.~Bl{\"u}mlein and J.~Brunner, ``{New Exclusion Limits on Dark Gauge Forces from Proton Bremsstrahlung in Beam-Dump Data},'' \href{http://dx.doi.org/10.1016/j.physletb.2014.02.029}{{\em Phys. Lett. B} {\bf 731} (2014)  320--326}, \href{http://arxiv.org/abs/1311.3870}{{\tt arXiv:1311.3870 [hep-ph]}}.

\bibitem{deNiverville:2016rqh}
P.~deNiverville, C.-Y. Chen, M.~Pospelov, and A.~Ritz, ``{Light dark matter in neutrino beams: production modelling and scattering signatures at MiniBooNE, T2K and SHiP},'' \href{http://dx.doi.org/10.1103/PhysRevD.95.035006}{{\em Phys. Rev. D} {\bf 95} (2017) no.~3, 035006}, \href{http://arxiv.org/abs/1609.01770}{{\tt arXiv:1609.01770 [hep-ph]}}.

\bibitem{Foroughi-Abari:2021zbm}
S.~Foroughi-Abari and A.~Ritz, ``{Dark sector production via proton bremsstrahlung},'' \href{http://dx.doi.org/10.1103/PhysRevD.105.095045}{{\em Phys. Rev. D} {\bf 105} (2022) no.~9, 095045}, \href{http://arxiv.org/abs/2108.05900}{{\tt arXiv:2108.05900 [hep-ph]}}.

\bibitem{Foroughi-Abari:2024xlj}
S.~Foroughi-Abari, P.~Reimitz, and A.~Ritz, ``{A Closer Look at Dark Vector Splitting Functions in Proton Bremsstrahlung},'' \href{http://arxiv.org/abs/2409.09123}{{\tt arXiv:2409.09123 [hep-ph]}}.

\bibitem{Gorbunov:2024vrc}
D.~Gorbunov and E.~Kriukova, ``{Pauli form factor contributions to the inelastic proton bremsstrahlung and dark photon production},'' \href{http://dx.doi.org/10.1007/JHEP02(2025)018}{{\em JHEP} {\bf 02} (2025)  018}, \href{http://arxiv.org/abs/2409.11386}{{\tt arXiv:2409.11386 [hep-ph]}}.

\bibitem{Gorbunov:2024iyu}
D.~Gorbunov and E.~Kriukova, ``{Dark Photon Production Via Inelastic Proton Bremsstrahlung with Pauli Form Factor},'' \href{http://dx.doi.org/10.1134/S1063779624701867}{{\em Phys. Part. Nucl.} {\bf 56} (2025) no.~2, 506--510}, \href{http://arxiv.org/abs/2409.11089}{{\tt arXiv:2409.11089 [hep-ph]}}.

\bibitem{Okubo:1963fa}
S.~Okubo, ``{Phi meson and unitary symmetry model},'' \href{http://dx.doi.org/10.1016/S0375-9601(63)92548-9}{{\em Phys. Lett.} {\bf 5} (1963)  165--168}.

\bibitem{Zweig:1964jf}
G.~Zweig, {\em {An SU(3) model for strong interaction symmetry and its breaking. Version 2}}, \href{http://dx.doi.org/10.17181/CERN-TH-412}{pp.~22--101}.
\newblock 2, 1964.

\bibitem{Iizuka:1966fk}
J.~Iizuka, ``{Systematics and phenomenology of meson family},'' \href{http://dx.doi.org/10.1143/PTPS.37.21}{{\em Prog. Theor. Phys. Suppl.} {\bf 37} (1966)  21--34}.

\bibitem{Czyz:2014sha}
H.~Czy{\.z}, J.~H. K{\"u}hn, and S.~Tracz, ``{Nucleon form factors and final state radiative corrections to $e^+e^- \to p\bar{p}\gamma$},'' \href{http://dx.doi.org/10.1103/PhysRevD.90.114021}{{\em Phys. Rev. D} {\bf 90} (2014) no.~11, 114021}, \href{http://arxiv.org/abs/1407.7995}{{\tt arXiv:1407.7995 [hep-ph]}}.

\bibitem{Lepage:1980fj}
G.~P. Lepage and S.~J. Brodsky, ``{Exclusive Processes in Perturbative Quantum Chromodynamics},'' \href{http://dx.doi.org/10.1103/PhysRevD.22.2157}{{\em Phys. Rev. D} {\bf 22} (1980)  2157}.

\bibitem{Adamuscin:2016rer}
C.~Adamu{\v{s}}{\v{c}}in, E.~Barto{\v{s}}, S.~Dubni{\v{c}}ka, and A.~Z. Dubni{\v{c}}kov{\'a}, ``{Numerical values of $f^F$, $f^D$, $f^S$ coupling constants in $SU(3)$ invariant interaction Lagrangian of vector-meson nonet with $1/2^+$ octet baryons},'' \href{http://dx.doi.org/10.1103/PhysRevC.93.055208}{{\em Phys. Rev. C} {\bf 93} (2016) no.~5, 055208}, \href{http://arxiv.org/abs/1601.06190}{{\tt arXiv:1601.06190 [hep-ph]}}.

\bibitem{BESIII:2015axk}
{\bf BESIII} Collaboration, M.~Ablikim {\em et al.}, ``{Measurement of the proton form factor by studying $e^{+} e^{-}\rightarrow p\bar{p}$},'' \href{http://dx.doi.org/10.1103/PhysRevD.91.112004}{{\em Phys. Rev. D} {\bf 91} (2015) no.~11, 112004}, \href{http://arxiv.org/abs/1504.02680}{{\tt arXiv:1504.02680 [hep-ex]}}.

\bibitem{Antonelli:1998fv}
A.~Antonelli {\em et al.}, ``{The first measurement of the neutron electromagnetic form-factors in the timelike region},'' \href{http://dx.doi.org/10.1016/S0550-3213(98)00083-2}{{\em Nucl. Phys. B} {\bf 517} (1998)  3--35}.

\bibitem{Delcourt:1979ed}
B.~Delcourt {\em et al.}, ``{Study of the Reaction $e^+ e^- \to p \bar{p}$ in the Total Energy Range 1925-{MeV} - 2180-{MeV}},'' \href{http://dx.doi.org/10.1016/0370-2693(79)90864-5}{{\em Phys. Lett. B} {\bf 86} (1979)  395--398}.

\bibitem{CLEO:2005tiu}
{\bf CLEO} Collaboration, T.~K. Pedlar {\em et al.}, ``{Precision measurements of the timelike electromagnetic form-factors of pion, kaon, and proton},'' \href{http://dx.doi.org/10.1103/PhysRevLett.95.261803}{{\em Phys. Rev. Lett.} {\bf 95} (2005)  261803}, \href{http://arxiv.org/abs/hep-ex/0510005}{{\tt arXiv:hep-ex/0510005}}.

\bibitem{Castellano:1973wh}
M.~Castellano, G.~Di~Giugno, J.~W. Humphrey, E.~Sassi~Palmieri, G.~Troise, U.~Troya, and S.~Vitale, ``{The reaction e+ e- ---{\ensuremath{>}} p anti-p at a total energy of 2.1 gev},'' \href{http://dx.doi.org/10.1007/BF02734600}{{\em Nuovo Cim. A} {\bf 14} (1973)  1--20}.

\bibitem{E760:1992rvj}
{\bf E760} Collaboration, T.~A. Armstrong {\em et al.}, ``{Measurement of the proton electromagnetic form-factors in the timelike region at $8.9-GeV^{2} - 13-GeV^{2}$},'' \href{http://dx.doi.org/10.1103/PhysRevLett.70.1212}{{\em Phys. Rev. Lett.} {\bf 70} (1993)  1212--1215}.

\bibitem{E835:1999mlt}
{\bf E835} Collaboration, M.~Ambrogiani {\em et al.}, ``{Measurements of the magnetic form-factor of the proton in the timelike region at large momentum transfer},'' \href{http://dx.doi.org/10.1103/PhysRevD.60.032002}{{\em Phys. Rev. D} {\bf 60} (1999)  032002}.

\bibitem{Andreotti:2003bt}
M.~Andreotti {\em et al.}, ``{Measurements of the Magnetic Form Factor of the Proton for Timelike Momentum Transfers},'' \href{http://dx.doi.org/10.1016/S0370-2693(03)00300-9}{{\em Phys. Lett. B} {\bf 559} (2003)  20--25}.

\bibitem{BaBar:2013ves}
{\bf BaBar} Collaboration, J.~P. Lees {\em et al.}, ``{Study of $e^+e^- \to p \bar{p}$ via initial-state radiation at BABAR},'' \href{http://dx.doi.org/10.1103/PhysRevD.87.092005}{{\em Phys. Rev. D} {\bf 87} (2013) no.~9, 092005}, \href{http://arxiv.org/abs/1302.0055}{{\tt arXiv:1302.0055 [hep-ex]}}.

\bibitem{Puckett:2011xg}
A.~J.~R. Puckett {\em et al.}, ``{Final Analysis of Proton Form Factor Ratio Data at $\mathbf{Q^2 = 4.0}$, 4.8 and 5.6 GeV$\mathbf{^2}$},'' \href{http://dx.doi.org/10.1103/PhysRevC.85.045203}{{\em Phys. Rev. C} {\bf 85} (2012)  045203}, \href{http://arxiv.org/abs/1102.5737}{{\tt arXiv:1102.5737 [nucl-ex]}}.

\bibitem{Puckett:2017flj}
A.~J.~R. Puckett {\em et al.}, ``{Polarization Transfer Observables in Elastic Electron Proton Scattering at $Q^2 = $2.5, 5.2, 6.8, and 8.5 GeV$^2$},'' \href{http://dx.doi.org/10.1103/PhysRevC.96.055203}{{\em Phys. Rev. C} {\bf 96} (2017) no.~5, 055203}, \href{http://arxiv.org/abs/1707.08587}{{\tt arXiv:1707.08587 [nucl-ex]}}. [Erratum: Phys.Rev.C 98, 019907 (2018)].

\bibitem{Punjabi:2005wq}
V.~Punjabi {\em et al.}, ``{Proton elastic form-factor ratios to Q**2 = 3.5-GeV**2 by polarization transfer},'' \href{http://dx.doi.org/10.1103/PhysRevC.71.055202}{{\em Phys. Rev. C} {\bf 71} (2005)  055202}, \href{http://arxiv.org/abs/nucl-ex/0501018}{{\tt arXiv:nucl-ex/0501018}}. [Erratum: Phys.Rev.C 71, 069902 (2005)].

\bibitem{E94110:2004lsx}
{\bf E94110} Collaboration, M.~E. Christy {\em et al.}, ``{Measurements of electron proton elastic cross-sections for 0.4 {\ensuremath{<}} Q**2 {\ensuremath{<}} 5.5 (GeV/c)**2},'' \href{http://dx.doi.org/10.1103/PhysRevC.70.015206}{{\em Phys. Rev. C} {\bf 70} (2004)  015206}, \href{http://arxiv.org/abs/nucl-ex/0401030}{{\tt arXiv:nucl-ex/0401030}}.

\bibitem{Bernauer:2010zga}
J.~C. Bernauer, {\em {Measurement of the elastic electron-proton cross section and separation of the electric and magnetic form factor in the Q$^{2}$ range from 0.004 to 1 (GeV/c)$^{2}$}}.
\newblock PhD thesis, Mainz U., Inst. Kernphys., 2010.

\bibitem{BESIII:2019hdp}
{\bf BESIII} Collaboration, M.~Ablikim {\em et al.}, ``{Measurement of proton electromagnetic form factors in $e^+e^- \to p\bar{p}$ in the energy region 2.00 - 3.08 GeV},'' \href{http://dx.doi.org/10.1103/PhysRevLett.124.042001}{{\em Phys. Rev. Lett.} {\bf 124} (2020) no.~4, 042001}, \href{http://arxiv.org/abs/1905.09001}{{\tt arXiv:1905.09001 [hep-ex]}}.

\bibitem{Hohler:1976ax}
G.~Hohler, E.~Pietarinen, I.~Sabba~Stefanescu, F.~Borkowski, G.~G. Simon, V.~H. Walther, and R.~D. Wendling, ``{Analysis of Electromagnetic Nucleon Form-Factors},'' \href{http://dx.doi.org/10.1016/0550-3213(76)90449-1}{{\em Nucl. Phys. B} {\bf 114} (1976)  505--534}.

\bibitem{Bartel:1973rf}
W.~Bartel, F.~W. Busser, W.~r. Dix, R.~Felst, D.~Harms, H.~Krehbiel, P.~E. Kuhlmann, J.~McElroy, J.~Meyer, and G.~Weber, ``{Measurement of proton and neutron electromagnetic form-factors at squared four momentum transfers up to 3-GeV/c$^2$},'' \href{http://dx.doi.org/10.1016/0550-3213(73)90594-4}{{\em Nucl. Phys. B} {\bf 58} (1973)  429--475}.

\bibitem{Janssens:1965kd}
T.~Janssens, R.~Hofstadter, E.~B. Hughes, and M.~R. Yearian, ``{Proton form factors from elastic electron-proton scattering},'' \href{http://dx.doi.org/10.1103/PhysRev.142.922}{{\em Phys. Rev.} {\bf 142} (1966)  922--931}.

\bibitem{BESIII:2021rqk}
{\bf BESIII} Collaboration, M.~Ablikim {\em et al.}, ``{Measurement of proton electromagnetic form factors in the time-like region using initial state radiation at BESIII},'' \href{http://dx.doi.org/10.1016/j.physletb.2021.136328}{{\em Phys. Lett. B} {\bf 817} (2021)  136328}, \href{http://arxiv.org/abs/2102.10337}{{\tt arXiv:2102.10337 [hep-ex]}}.

\bibitem{JeffersonLaboratoryE93-038:2005ryd}
{\bf Jefferson Laboratory E93-038} Collaboration, B.~Plaster {\em et al.}, ``{Measurements of the neutron electric to magnetic form-factor ratio G(En) / G(Mn) via the H-2(polarized-e, e-prime,polarized-n)H-1 reaction to Q**2 = 1.45-(GeV/c)**2},'' \href{http://dx.doi.org/10.1103/PhysRevC.73.025205}{{\em Phys. Rev. C} {\bf 73} (2006)  025205}, \href{http://arxiv.org/abs/nucl-ex/0511025}{{\tt arXiv:nucl-ex/0511025}}.

\bibitem{BLAST:2008bub}
{\bf BLAST} Collaboration, E.~Geis {\em et al.}, ``{The Charge Form Factor of the Neutron at Low Momentum Transfer from the H-2-polarized (e-polarized, e-prime n) p Reaction},'' \href{http://dx.doi.org/10.1103/PhysRevLett.101.042501}{{\em Phys. Rev. Lett.} {\bf 101} (2008)  042501}, \href{http://arxiv.org/abs/0803.3827}{{\tt arXiv:0803.3827 [nucl-ex]}}.

\bibitem{Golak:2000nt}
J.~Golak, G.~Ziemer, H.~Kamada, H.~Witala, and W.~Gloeckle, ``{Extraction of electromagnetic neutron form-factors through inclusive and exclusive polarized electron scattering on polarized He-3 target},'' \href{http://dx.doi.org/10.1103/PhysRevC.63.034006}{{\em Phys. Rev. C} {\bf 63} (2001)  034006}, \href{http://arxiv.org/abs/nucl-th/0008008}{{\tt arXiv:nucl-th/0008008}}.

\bibitem{Glazier:2004ny}
D.~I. Glazier {\em et al.}, ``{Measurement of the electric form-factor of the neutron at Q**2 = 0.3-(GeV/c)**2 to 0.8-(GeV/c)**2},'' \href{http://dx.doi.org/10.1140/epja/i2004-10115-8}{{\em Eur. Phys. J. A} {\bf 24} (2005)  101--109}, \href{http://arxiv.org/abs/nucl-ex/0410026}{{\tt arXiv:nucl-ex/0410026}}.

\bibitem{JeffersonLabE93-026:2003tty}
{\bf Jefferson Lab E93-026} Collaboration, G.~Warren {\em et al.}, ``{Measurement of the electric form-factor of the neutron at $Q^2$ = 0.5 and 1.0 $GeV^2/c^2$},'' \href{http://dx.doi.org/10.1103/PhysRevLett.92.042301}{{\em Phys. Rev. Lett.} {\bf 92} (2004)  042301}, \href{http://arxiv.org/abs/nucl-ex/0308021}{{\tt arXiv:nucl-ex/0308021}}.

\bibitem{E93026:2001css}
{\bf E93026} Collaboration, H.~Zhu {\em et al.}, ``{A Measurement of the electric form-factor of the neutron through polarized-d (polarized-e, e-prime n)p at Q**2 = 0.5-(GeV/c)**2},'' \href{http://dx.doi.org/10.1103/PhysRevLett.87.081801}{{\em Phys. Rev. Lett.} {\bf 87} (2001)  081801}, \href{http://arxiv.org/abs/nucl-ex/0105001}{{\tt arXiv:nucl-ex/0105001}}.

\bibitem{Herberg:1999ud}
C.~Herberg {\em et al.}, ``{Determination of the neutron electric form-factor in the D(e,e' n)p reaction and the influence of nuclear binding},'' \href{http://dx.doi.org/10.1007/s100500050268}{{\em Eur. Phys. J. A} {\bf 5} (1999)  131--135}.

\bibitem{Ostrick:1999xa}
M.~Ostrick {\em et al.}, ``{Measurement of the neutron electric form-factor G(E,n) in the quasifree H-2(e(pol.),e' n(pol.))p reaction},'' \href{http://dx.doi.org/10.1103/PhysRevLett.83.276}{{\em Phys. Rev. Lett.} {\bf 83} (1999)  276--279}.

\bibitem{Passchier:1999cj}
I.~Passchier {\em et al.}, ``{The Charge form-factor of the neutron from the reaction polarized H-2(polarized e, e-prime n) p},'' \href{http://dx.doi.org/10.1103/PhysRevLett.82.4988}{{\em Phys. Rev. Lett.} {\bf 82} (1999)  4988--4991}, \href{http://arxiv.org/abs/nucl-ex/9907012}{{\tt arXiv:nucl-ex/9907012}}.

\bibitem{Eden:1994ji}
T.~Eden {\em et al.}, ``{Electric form factor of the neutron from the $^{2}H(\overrightarrow{e},e’\overrightarrow{n})^{1}H$ reaction at $Q^{2} =$ 0.255 (GeV/c)$^2$},'' \href{http://dx.doi.org/10.1103/PhysRevC.50.R1749}{{\em Phys. Rev. C} {\bf 50} (1994) no.~4, R1749--R1753}.

\bibitem{Berard:1973xrv}
R.~W. Berard, F.~R. Buskirk, E.~B. Dally, J.~N. Dyer, X.~K. Maruyama, R.~L. Topping, and T.~J. Traverso, ``{Elastic electron deuteron scattering},'' \href{http://dx.doi.org/10.1016/0370-2693(73)90622-9}{{\em Phys. Lett. B} {\bf 47} (1973)  355--358}.

\bibitem{Rock:1982gf}
S.~Rock, R.~G. Arnold, P.~E. Bosted, B.~T. Chertok, B.~A. Mecking, I.~A. Schmidt, Z.~M. Szalata, R.~York, and R.~Zdarko, ``{Measurement of Elastic electron - Neutron Cross-Sections Up to Q**2 = 10-(GeV/c)**2},'' \href{http://dx.doi.org/10.1103/PhysRevLett.49.1139}{{\em Phys. Rev. Lett.} {\bf 49} (1982)  1139}.

\bibitem{Lung:1992bu}
A.~Lung {\em et al.}, ``{Measurements of the electric and magnetic form-factors of the neutron from Q**2 = 1.75-GeV/c**2 to 4-GeV/c**2},'' \href{http://dx.doi.org/10.1103/PhysRevLett.70.718}{{\em Phys. Rev. Lett.} {\bf 70} (1993)  718--721}.

\bibitem{JeffersonLabE95-001:2006dax}
{\bf Jefferson Lab E95-001} Collaboration, B.~Anderson {\em et al.}, ``{Extraction of the Neutron Magnetic Form Factor from Quasi-elastic $^{3}\vec{He}(\vec{e},e')$ at Q$^2$ = 0.1 - 0.6 (GeV/c)$^2$},'' \href{http://dx.doi.org/10.1103/PhysRevC.75.034003}{{\em Phys. Rev. C} {\bf 75} (2007)  034003}, \href{http://arxiv.org/abs/nucl-ex/0605006}{{\tt arXiv:nucl-ex/0605006}}.

\bibitem{Kubon:2001rj}
G.~Kubon {\em et al.}, ``{Precise neutron magnetic form-factors},'' \href{http://dx.doi.org/10.1016/S0370-2693(01)01386-7}{{\em Phys. Lett. B} {\bf 524} (2002)  26--32}, \href{http://arxiv.org/abs/nucl-ex/0107016}{{\tt arXiv:nucl-ex/0107016}}.

\bibitem{Anklin:1998ae}
H.~Anklin {\em et al.}, ``{Precise measurements of the neutron magnetic form-factor},'' \href{http://dx.doi.org/10.1016/S0370-2693(98)00442-0}{{\em Phys. Lett. B} {\bf 428} (1998)  248--253}.

\bibitem{BESIII:2021tbq}
{\bf BESIII} Collaboration, M.~Ablikim {\em et al.}, ``{Oscillating features in the electromagnetic structure of the neutron},'' \href{http://dx.doi.org/10.1038/s41567-021-01345-6}{{\em Nature Phys.} {\bf 17} (2021) no.~11, 1200--1204}, \href{http://arxiv.org/abs/2103.12486}{{\tt arXiv:2103.12486 [hep-ex]}}.

\bibitem{Druzhinin:2020fgn}
V.~Druzhinin {\em et al.}, ``{Study of $e^+e^-$ annihilation into hadrons with the SND detector at the VEPP-2000 collider},'' \href{http://dx.doi.org/10.22323/1.364.0509}{{\em PoS} {\bf EPS-HEP2019} (2020)  509}.

\bibitem{Hogg:2010xxx}
D.~W. Hogg, J.~Bovy, and D.~Lang, ``Data analysis recipes: Fitting a model to data,'' \href{http://arxiv.org/abs/1008.4686}{{\tt arXiv:1008.4686 [astro-ph.IM]}}. \url{https://arxiv.org/abs/1008.4686}.

\bibitem{ParticleDataGroup:2024cfk}
{\bf Particle Data Group} Collaboration, S.~Navas {\em et al.}, ``{Review of particle physics},'' \href{http://dx.doi.org/10.1103/PhysRevD.110.030001}{{\em Phys. Rev. D} {\bf 110} (2024) no.~3, 030001}.

\bibitem{2020SciPy-NMeth}
P.~Virtanen {\em et al.}, ``{{SciPy} 1.0: Fundamenta Algorithms for Scientific Computing in Python},'' \href{http://dx.doi.org/10.1038/s41592-019-0686-2}{{\em Nature Methods} {\bf 17} (2020)  261--272}.

\bibitem{James:1975dr}
F.~James and M.~Roos, ``{Minuit: A System for Function Minimization and Analysis of the Parameter Errors and Correlations},'' \href{http://dx.doi.org/10.1016/0010-4655(75)90039-9}{{\em Comput. Phys. Commun.} {\bf 10} (1975)  343--367}.

\bibitem{iminuit}
H.~Dembinski and P.~O. et~al., ``scikit-hep/iminuit,''. \url{https://doi.org/10.5281/zenodo.3949207}.

\bibitem{Likhoded:2010pc}
A.~K. Likhoded, A.~V. Luchinsky, and A.~A. Novoselov, ``{Light hadron production in inclusive pp-scattering at LHC},'' \href{http://dx.doi.org/10.1103/PhysRevD.82.114006}{{\em Phys. Rev. D} {\bf 82} (2010)  114006}, \href{http://arxiv.org/abs/1005.1827}{{\tt arXiv:1005.1827 [hep-ph]}}.

\bibitem{FASER:2022hcn}
{\bf FASER} Collaboration, H.~Abreu {\em et al.}, ``{The FASER detector},'' \href{http://dx.doi.org/10.1088/1748-0221/19/05/P05066}{{\em JINST} {\bf 19} (2024) no.~05, P05066}, \href{http://arxiv.org/abs/2207.11427}{{\tt arXiv:2207.11427 [physics.ins-det]}}.

\bibitem{FASER:2019dxq}
{\bf FASER} Collaboration, H.~Abreu {\em et al.}, ``{Detecting and Studying High-Energy Collider Neutrinos with FASER at the LHC},'' \href{http://dx.doi.org/10.1140/epjc/s10052-020-7631-5}{{\em Eur. Phys. J. C} {\bf 80} (2020) no.~1, 61}, \href{http://arxiv.org/abs/1908.02310}{{\tt arXiv:1908.02310 [hep-ex]}}.

\bibitem{FASER:2020gpr}
{\bf FASER} Collaboration, H.~Abreu {\em et al.}, ``{Technical Proposal: FASERnu},'' \href{http://arxiv.org/abs/2001.03073}{{\tt arXiv:2001.03073 [physics.ins-det]}}.

\bibitem{FASER:2025qaf}
{\bf FASER} Collaboration, R.~Mammen~Abraham {\em et al.}, ``{Reconstruction and Performance Evaluation of FASER's Emulsion Detector at the LHC},'' \href{http://arxiv.org/abs/2504.13008}{{\tt arXiv:2504.13008 [physics.ins-det]}}.

\bibitem{FASER:2024hoe}
{\bf FASER} Collaboration, R.~Mammen~Abraham {\em et al.}, ``{First Measurement of {\ensuremath{\nu}}e and {\ensuremath{\nu}}{\ensuremath{\mu}} Interaction Cross Sections at the LHC with FASER{\textquoteright}s Emulsion Detector},'' \href{http://dx.doi.org/10.1103/PhysRevLett.133.021802}{{\em Phys. Rev. Lett.} {\bf 133} (2024) no.~2, 021802}, \href{http://arxiv.org/abs/2403.12520}{{\tt arXiv:2403.12520 [hep-ex]}}.

\bibitem{FASER:2018ceo}
{\bf FASER} Collaboration, A.~Ariga {\em et al.}, ``{Letter of Intent for FASER: ForwArd Search ExpeRiment at the LHC},'' \href{http://arxiv.org/abs/1811.10243}{{\tt arXiv:1811.10243 [physics.ins-det]}}.

\bibitem{FASER:2018bac}
{\bf FASER} Collaboration, A.~Ariga {\em et al.}, ``{Technical Proposal for FASER: ForwArd Search ExpeRiment at the LHC},'' \href{http://arxiv.org/abs/1812.09139}{{\tt arXiv:1812.09139 [physics.ins-det]}}.

\bibitem{FASER:2018eoc}
{\bf FASER} Collaboration, A.~Ariga {\em et al.}, ``{FASER's physics reach for long-lived particles},'' \href{http://dx.doi.org/10.1103/PhysRevD.99.095011}{{\em Phys. Rev. D} {\bf 99} (2019) no.~9, 095011}, \href{http://arxiv.org/abs/1811.12522}{{\tt arXiv:1811.12522 [hep-ph]}}.

\bibitem{FASER:2019aik}
{\bf FASER} Collaboration, A.~Ariga {\em et al.}, ``{FASER: ForwArd Search ExpeRiment at the LHC},'' \href{http://arxiv.org/abs/1901.04468}{{\tt arXiv:1901.04468 [hep-ex]}}.

\bibitem{FASER:2021ljd}
{\bf FASER} Collaboration, H.~Abreu {\em et al.}, ``{The tracking detector of the FASER experiment},'' \href{http://dx.doi.org/10.1016/j.nima.2022.166825}{{\em Nucl. Instrum. Meth. A} {\bf 1034} (2022)  166825}, \href{http://arxiv.org/abs/2112.01116}{{\tt arXiv:2112.01116 [physics.ins-det]}}.

\bibitem{FASER:2023tle}
{\bf FASER} Collaboration, H.~Abreu {\em et al.}, ``{Search for dark photons with the FASER detector at the LHC},'' \href{http://dx.doi.org/10.1016/j.physletb.2023.138378}{{\em Phys. Lett. B} {\bf 848} (2024)  138378}, \href{http://arxiv.org/abs/2308.05587}{{\tt arXiv:2308.05587 [hep-ex]}}.

\bibitem{FASER:2024bbl}
{\bf FASER} Collaboration, R.~Mammen~Abraham {\em et al.}, ``{Shining light on the dark sector: search for axion-like particles and other new physics in photonic final states with FASER},'' \href{http://dx.doi.org/10.1007/JHEP01(2025)199}{{\em JHEP} {\bf 01} (2025)  199}, \href{http://arxiv.org/abs/2410.10363}{{\tt arXiv:2410.10363 [hep-ex]}}.

\bibitem{LHCf:2018gbv}
{\bf LHCf} Collaboration, O.~Adriani {\em et al.}, ``{Measurement of inclusive forward neutron production cross section in proton-proton collisions at $ \sqrt{s}=13 $ TeV with the LHCf Arm2 detector},'' \href{http://dx.doi.org/10.1007/JHEP11(2018)073}{{\em JHEP} {\bf 11} (2018)  073}, \href{http://arxiv.org/abs/1808.09877}{{\tt arXiv:1808.09877 [hep-ex]}}.

\bibitem{Pierog:2013ria}
T.~Pierog, I.~Karpenko, J.~M. Katzy, E.~Yatsenko, and K.~Werner, ``{EPOS LHC: Test of collective hadronization with data measured at the CERN Large Hadron Collider},'' \href{http://dx.doi.org/10.1103/PhysRevC.92.034906}{{\em Phys. Rev. C} {\bf 92} (2015) no.~3, 034906}, \href{http://arxiv.org/abs/1306.0121}{{\tt arXiv:1306.0121 [hep-ph]}}.

\bibitem{Riehn:2019jet}
F.~Riehn, R.~Engel, A.~Fedynitch, T.~K. Gaisser, and T.~Stanev, ``{Hadronic interaction model Sibyll 2.3d and extensive air showers},'' \href{http://dx.doi.org/10.1103/PhysRevD.102.063002}{{\em Phys. Rev. D} {\bf 102} (2020) no.~6, 063002}, \href{http://arxiv.org/abs/1912.03300}{{\tt arXiv:1912.03300 [hep-ph]}}.

\bibitem{Ostapchenko:2010vb}
S.~Ostapchenko, ``{Monte Carlo treatment of hadronic interactions in enhanced Pomeron scheme: I. QGSJET-II model},'' \href{http://dx.doi.org/10.1103/PhysRevD.83.014018}{{\em Phys. Rev. D} {\bf 83} (2011)  014018}, \href{http://arxiv.org/abs/1010.1869}{{\tt arXiv:1010.1869 [hep-ph]}}.

\bibitem{Kling:2021fwx}
F.~Kling and S.~Trojanowski, ``{Forward experiment sensitivity estimator for the LHC and future hadron colliders},'' \href{http://dx.doi.org/10.1103/PhysRevD.104.035012}{{\em Phys. Rev. D} {\bf 104} (2021) no.~3, 035012}, \href{http://arxiv.org/abs/2105.07077}{{\tt arXiv:2105.07077 [hep-ph]}}.

\bibitem{Heinrich:2021gyp}
L.~Heinrich, M.~Feickert, G.~Stark, and K.~Cranmer, ``{pyhf: pure-Python implementation of HistFactory statistical models},'' \href{http://dx.doi.org/10.21105/joss.02823}{{\em J. Open Source Softw.} {\bf 6} (2021) no.~58, 2823}.

\bibitem{Merkel:2014avp}
H.~Merkel {\em et al.}, ``{Search at the Mainz Microtron for Light Massive Gauge Bosons Relevant for the Muon g-2 Anomaly},'' \href{http://dx.doi.org/10.1103/PhysRevLett.112.221802}{{\em Phys. Rev. Lett.} {\bf 112} (2014) no.~22, 221802}, \href{http://arxiv.org/abs/1404.5502}{{\tt arXiv:1404.5502 [hep-ex]}}.

\bibitem{BaBar:2014zli}
{\bf BaBar} Collaboration, J.~P. Lees {\em et al.}, ``{Search for a Dark Photon in $e^+e^-$ Collisions at BaBar},'' \href{http://dx.doi.org/10.1103/PhysRevLett.113.201801}{{\em Phys. Rev. Lett.} {\bf 113} (2014) no.~20, 201801}, \href{http://arxiv.org/abs/1406.2980}{{\tt arXiv:1406.2980 [hep-ex]}}.

\bibitem{NA482:2015wmo}
{\bf NA48/2} Collaboration, J.~R. Batley {\em et al.}, ``{Search for the dark photon in $\pi^0$ decays},'' \href{http://dx.doi.org/10.1016/j.physletb.2015.04.068}{{\em Phys. Lett. B} {\bf 746} (2015)  178--185}, \href{http://arxiv.org/abs/1504.00607}{{\tt arXiv:1504.00607 [hep-ex]}}.

\bibitem{NA64:2019auh}
{\bf NA64} Collaboration, D.~Banerjee {\em et al.}, ``{Improved limits on a hypothetical X(16.7) boson and a dark photon decaying into $e^+e^-$ pairs},'' \href{http://dx.doi.org/10.1103/PhysRevD.101.071101}{{\em Phys. Rev. D} {\bf 101} (2020) no.~7, 071101}, \href{http://arxiv.org/abs/1912.11389}{{\tt arXiv:1912.11389 [hep-ex]}}.

\bibitem{NA62:2023nhs}
{\bf NA62} Collaboration, E.~Cortina~Gil {\em et al.}, ``{Search for Leptonic Decays of Dark Photons at NA62},'' \href{http://dx.doi.org/10.1103/PhysRevLett.133.111802}{{\em Phys. Rev. Lett.} {\bf 133} (2024) no.~11, 111802}, \href{http://arxiv.org/abs/2312.12055}{{\tt arXiv:2312.12055 [hep-ex]}}.

\bibitem{Foguel:2022ppx}
A.~L. Foguel, P.~Reimitz, and R.~Z. Funchal, ``{A robust description of hadronic decays in light vector mediator models},'' \href{http://dx.doi.org/10.1007/JHEP04(2022)119}{{\em JHEP} {\bf 04} (2022)  119}, \href{http://arxiv.org/abs/2201.01788}{{\tt arXiv:2201.01788 [hep-ph]}}.

\bibitem{CHARM:1985anb}
{\bf CHARM} Collaboration, F.~Bergsma {\em et al.}, ``{Search for Axion Like Particle Production in 400-{GeV} Proton - Copper Interactions},'' \href{http://dx.doi.org/10.1016/0370-2693(85)90400-9}{{\em Phys. Lett. B} {\bf 157} (1985)  458--462}.

\bibitem{Davier:1989wz}
M.~Davier and H.~Nguyen~Ngoc, ``{An Unambiguous Search for a Light Higgs Boson},'' \href{http://dx.doi.org/10.1016/0370-2693(89)90174-3}{{\em Phys. Lett. B} {\bf 229} (1989)  150--155}.

\bibitem{Blumlein:1991xh}
J.~Blumlein {\em et al.}, ``{Limits on the mass of light (pseudo)scalar particles from Bethe-Heitler e+ e- and mu+ mu- pair production in a proton - iron beam dump experiment},'' \href{http://dx.doi.org/10.1142/S0217751X9200171X}{{\em Int. J. Mod. Phys. A} {\bf 7} (1992)  3835--3850}.

\bibitem{Ilten:2018crw}
P.~Ilten, Y.~Soreq, M.~Williams, and W.~Xue, ``{Serendipity in dark photon searches},'' \href{http://dx.doi.org/10.1007/JHEP06(2018)004}{{\em JHEP} {\bf 06} (2018)  004}, \href{http://arxiv.org/abs/1801.04847}{{\tt arXiv:1801.04847 [hep-ph]}}.

\bibitem{CHARM-II:1993phx}
{\bf CHARM-II} Collaboration, P.~Vilain {\em et al.}, ``{Measurement of differential cross-sections for muon-neutrino electron scattering},'' \href{http://dx.doi.org/10.1016/0370-2693(93)90408-A}{{\em Phys. Lett. B} {\bf 302} (1993)  351--355}.

\bibitem{Bilmis:2015lja}
S.~Bilmis, I.~Turan, T.~M. Aliev, M.~Deniz, L.~Singh, and H.~T. Wong, ``{Constraints on Dark Photon from Neutrino-Electron Scattering Experiments},'' \href{http://dx.doi.org/10.1103/PhysRevD.92.033009}{{\em Phys. Rev. D} {\bf 92} (2015) no.~3, 033009}, \href{http://arxiv.org/abs/1502.07763}{{\tt arXiv:1502.07763 [hep-ph]}}.

\bibitem{NA64:2022yly}
{\bf NA64} Collaboration, Y.~M. Andreev {\em et al.}, ``{Search for a New B-L Z' Gauge Boson with the NA64 Experiment at CERN},'' \href{http://dx.doi.org/10.1103/PhysRevLett.129.161801}{{\em Phys. Rev. Lett.} {\bf 129} (2022) no.~16, 161801}, \href{http://arxiv.org/abs/2207.09979}{{\tt arXiv:2207.09979 [hep-ex]}}.

\bibitem{Blumlein:1990ay}
J.~Blumlein {\em et al.}, ``{Limits on neutral light scalar and pseudoscalar particles in a proton beam dump experiment},'' \href{http://dx.doi.org/10.1007/BF01548556}{{\em Z. Phys. C} {\bf 51} (1991)  341--350}.

\bibitem{Anastasi:2015qla}
A.~Anastasi {\em et al.}, ``{Limit on the production of a low-mass vector boson in $\mathrm{e}^{+}\mathrm{e}^{-} \to \mathrm{U}\gamma$, $\mathrm{U} \to \mathrm{e}^{+}\mathrm{e}^{-}$ with the KLOE experiment},'' \href{http://dx.doi.org/10.1016/j.physletb.2015.10.003}{{\em Phys. Lett. B} {\bf 750} (2015)  633--637}, \href{http://arxiv.org/abs/1509.00740}{{\tt arXiv:1509.00740 [hep-ex]}}.

\bibitem{KLOE-2:2012lii}
{\bf KLOE-2} Collaboration, D.~Babusci {\em et al.}, ``{Limit on the production of a light vector gauge boson in phi meson decays with the KLOE detector},'' \href{http://dx.doi.org/10.1016/j.physletb.2013.01.067}{{\em Phys. Lett. B} {\bf 720} (2013)  111--115}, \href{http://arxiv.org/abs/1210.3927}{{\tt arXiv:1210.3927 [hep-ex]}}.

\bibitem{LHCb:2017trq}
{\bf LHCb} Collaboration, R.~Aaij {\em et al.}, ``{Search for Dark Photons Produced in 13 TeV $pp$ Collisions},'' \href{http://dx.doi.org/10.1103/PhysRevLett.120.061801}{{\em Phys. Rev. Lett.} {\bf 120} (2018) no.~6, 061801}, \href{http://arxiv.org/abs/1710.02867}{{\tt arXiv:1710.02867 [hep-ex]}}.

\bibitem{LHCb:2019vmc}
{\bf LHCb} Collaboration, R.~Aaij {\em et al.}, ``{Search for $A'\to\mu^+\mu^-$ Decays},'' \href{http://dx.doi.org/10.1103/PhysRevLett.124.041801}{{\em Phys. Rev. Lett.} {\bf 124} (2020) no.~4, 041801}, \href{http://arxiv.org/abs/1910.06926}{{\tt arXiv:1910.06926 [hep-ex]}}.

\bibitem{Dror:2017ehi}
J.~A. Dror, R.~Lasenby, and M.~Pospelov, ``{New constraints on light vectors coupled to anomalous currents},'' \href{http://dx.doi.org/10.1103/PhysRevLett.119.141803}{{\em Phys. Rev. Lett.} {\bf 119} (2017) no.~14, 141803}, \href{http://arxiv.org/abs/1705.06726}{{\tt arXiv:1705.06726 [hep-ph]}}.

\bibitem{Feng:2016ysn}
J.~L. Feng, B.~Fornal, I.~Galon, S.~Gardner, J.~Smolinsky, T.~M.~P. Tait, and P.~Tanedo, ``{Particle physics models for the 17 MeV anomaly in beryllium nuclear decays},'' \href{http://dx.doi.org/10.1103/PhysRevD.95.035017}{{\em Phys. Rev. D} {\bf 95} (2017) no.~3, 035017}, \href{http://arxiv.org/abs/1608.03591}{{\tt arXiv:1608.03591 [hep-ph]}}.

\bibitem{Feng:2020mbt}
J.~L. Feng, T.~M.~P. Tait, and C.~B. Verhaaren, ``{Dynamical Evidence For a Fifth Force Explanation of the ATOMKI Nuclear Anomalies},'' \href{http://dx.doi.org/10.1103/PhysRevD.102.036016}{{\em Phys. Rev. D} {\bf 102} (2020) no.~3, 036016}, \href{http://arxiv.org/abs/2006.01151}{{\tt arXiv:2006.01151 [hep-ph]}}.

\bibitem{Krasznahorkay:2015iga}
A.~J. Krasznahorkay {\em et al.}, ``{Observation of Anomalous Internal Pair Creation in Be8 : A Possible Indication of a Light, Neutral Boson},'' \href{http://dx.doi.org/10.1103/PhysRevLett.116.042501}{{\em Phys. Rev. Lett.} {\bf 116} (2016) no.~4, 042501}, \href{http://arxiv.org/abs/1504.01527}{{\tt arXiv:1504.01527 [nucl-ex]}}.

\bibitem{Firak:2020eil}
D.~S. Firak {\em et al.}, ``{Confirmation of the existence of the X17 particle},'' \href{http://dx.doi.org/10.1051/epjconf/202023204005}{{\em EPJ Web Conf.} {\bf 232} (2020)  04005}.

\bibitem{Alves:2023ree}
D.~S.~M. Alves {\em et al.}, ``{Shedding light on X17: community report},'' \href{http://dx.doi.org/10.1140/epjc/s10052-023-11271-x}{{\em Eur. Phys. J. C} {\bf 83} (2023) no.~3, 230}.

\bibitem{Dev:2023zts}
P.~S.~B. Dev, B.~Dutta, T.~Han, A.~Karthikeyan, D.~Kim, and H.~Kim, ``{New physics at a neutron beam dump},'' \href{http://dx.doi.org/10.1103/PhysRevD.110.L051703}{{\em Phys. Rev. D} {\bf 110} (2024) no.~5, L051703}, \href{http://arxiv.org/abs/2311.10078}{{\tt arXiv:2311.10078 [hep-ph]}}.

\bibitem{KLOE-2:2018kqf}
{\bf KLOE-2} Collaboration, A.~Anastasi {\em et al.}, ``{Combined limit on the production of a light gauge boson decaying into $\mu^+\mu^-$ and $\pi^+\pi^-$},'' \href{http://dx.doi.org/10.1016/j.physletb.2018.08.012}{{\em Phys. Lett. B} {\bf 784} (2018)  336--341}, \href{http://arxiv.org/abs/1807.02691}{{\tt arXiv:1807.02691 [hep-ex]}}.

\bibitem{Bossi:2025ptv}
F.~Bossi {\em et al.}, ``{Search for a new 17 MeV resonance via $e^+e^-$ annihilation with the PADME Experiment},'' \href{http://arxiv.org/abs/2505.24797}{{\tt arXiv:2505.24797 [hep-ex]}}.

\bibitem{Riordan:1987aw}
E.~M. Riordan {\em et al.}, ``{A Search for Short Lived Axions in an Electron Beam Dump Experiment},'' \href{http://dx.doi.org/10.1103/PhysRevLett.59.755}{{\em Phys. Rev. Lett.} {\bf 59} (1987)  755}.

\bibitem{Arias-Aragon:2025wdt}
F.~Arias-Arag{\'o}n, G.~G. di~Cortona, E.~Nardi, and C.~Toni, ``{Combined Evidence for the $X_{17}$ Boson After PADME Results on Resonant Production in Positron Annihilation},'' \href{http://arxiv.org/abs/2504.11439}{{\tt arXiv:2504.11439 [hep-ph]}}.

\bibitem{Holdom:1985ag}
B.~Holdom, ``{Two U(1)'s and Epsilon Charge Shifts},'' \href{http://dx.doi.org/10.1016/0370-2693(86)91377-8}{{\em Phys. Lett. B} {\bf 166} (1986)  196--198}.

\bibitem{Foroughi-Abari:2020qar}
S.~Foroughi-Abari, F.~Kling, and Y.-D. Tsai, ``{Looking forward to millicharged dark sectors at the LHC},'' \href{http://dx.doi.org/10.1103/PhysRevD.104.035014}{{\em Phys. Rev. D} {\bf 104} (2021) no.~3, 035014}, \href{http://arxiv.org/abs/2010.07941}{{\tt arXiv:2010.07941 [hep-ph]}}.

\bibitem{Haas:2014dda}
A.~Haas, C.~S. Hill, E.~Izaguirre, and I.~Yavin, ``{Looking for milli-charged particles with a new experiment at the LHC},'' \href{http://dx.doi.org/10.1016/j.physletb.2015.04.062}{{\em Phys. Lett. B} {\bf 746} (2015)  117--120}, \href{http://arxiv.org/abs/1410.6816}{{\tt arXiv:1410.6816 [hep-ph]}}.

\bibitem{Ball:2016zrp}
A.~Ball {\em et al.}, ``{A Letter of Intent to Install a milli-charged Particle Detector at LHC P5},'' \href{http://arxiv.org/abs/1607.04669}{{\tt arXiv:1607.04669 [physics.ins-det]}}.

\bibitem{milliQan:2021lne}
{\bf milliQan} Collaboration, A.~Ball {\em et al.}, ``{Sensitivity to millicharged particles in future proton-proton collisions at the LHC with the milliQan detector},'' \href{http://dx.doi.org/10.1103/PhysRevD.104.032002}{{\em Phys. Rev. D} {\bf 104} (2021) no.~3, 032002}, \href{http://arxiv.org/abs/2104.07151}{{\tt arXiv:2104.07151 [hep-ex]}}.

\bibitem{Gninenko:2018ter}
S.~N. Gninenko, D.~V. Kirpichnikov, and N.~V. Krasnikov, ``{Probing millicharged particles with NA64 experiment at CERN},'' \href{http://dx.doi.org/10.1103/PhysRevD.100.035003}{{\em Phys. Rev. D} {\bf 100} (2019) no.~3, 035003}, \href{http://arxiv.org/abs/1810.06856}{{\tt arXiv:1810.06856 [hep-ph]}}.

\bibitem{Marocco:2020dqu}
G.~Marocco and S.~Sarkar, ``{Blast from the past: Constraints on the dark sector from the BEBC WA66 beam dump experiment},'' \href{http://dx.doi.org/10.21468/SciPostPhys.10.2.043}{{\em SciPost Phys.} {\bf 10} (2021) no.~2, 043}, \href{http://arxiv.org/abs/2011.08153}{{\tt arXiv:2011.08153 [hep-ph]}}.

\bibitem{Prinz:1998ua}
A.~A. Prinz {\em et al.}, ``{Search for millicharged particles at SLAC},'' \href{http://dx.doi.org/10.1103/PhysRevLett.81.1175}{{\em Phys. Rev. Lett.} {\bf 81} (1998)  1175--1178}, \href{http://arxiv.org/abs/hep-ex/9804008}{{\tt arXiv:hep-ex/9804008}}.

\bibitem{Davidson:2000hf}
S.~Davidson, S.~Hannestad, and G.~Raffelt, ``{Updated bounds on millicharged particles},'' \href{http://dx.doi.org/10.1088/1126-6708/2000/05/003}{{\em JHEP} {\bf 05} (2000)  003}, \href{http://arxiv.org/abs/hep-ph/0001179}{{\tt arXiv:hep-ph/0001179}}.

\bibitem{OPAL:1995uwx}
{\bf OPAL} Collaboration, R.~Akers {\em et al.}, ``{Search for heavy charged particles and for particles with anomalous charge in $e^{+} e^{-}$ collisions at LEP},'' \href{http://dx.doi.org/10.1007/BF01571281}{{\em Z. Phys. C} {\bf 67} (1995)  203--212}.

\bibitem{CMS:2012xi}
{\bf CMS} Collaboration, S.~Chatrchyan {\em et al.}, ``{Search for Fractionally Charged Particles in $pp$ Collisions at $\sqrt{s}=7$ TeV},'' \href{http://dx.doi.org/10.1103/PhysRevD.87.092008}{{\em Phys. Rev. D} {\bf 87} (2013) no.~9, 092008}, \href{http://arxiv.org/abs/1210.2311}{{\tt arXiv:1210.2311 [hep-ex]}}. [Erratum: Phys.Rev.D 106, 099903 (2022)].

\bibitem{Jaeckel:2012yz}
J.~Jaeckel, M.~Jankowiak, and M.~Spannowsky, ``{LHC probes the hidden sector},'' \href{http://dx.doi.org/10.1016/j.dark.2013.06.001}{{\em Phys. Dark Univ.} {\bf 2} (2013)  111--117}, \href{http://arxiv.org/abs/1212.3620}{{\tt arXiv:1212.3620 [hep-ph]}}.

\bibitem{CMS:2024eyx}
{\bf CMS} Collaboration, A.~Hayrapetyan {\em et al.}, ``{Search for Fractionally Charged Particles in Proton-Proton Collisions at s=13{\,}{\,}TeV},'' \href{http://dx.doi.org/10.1103/PhysRevLett.134.131802}{{\em Phys. Rev. Lett.} {\bf 134} (2025) no.~13, 131802}, \href{http://arxiv.org/abs/2402.09932}{{\tt arXiv:2402.09932 [hep-ex]}}.

\bibitem{Magill:2018tbb}
G.~Magill, R.~Plestid, M.~Pospelov, and Y.-D. Tsai, ``{Millicharged particles in neutrino experiments},'' \href{http://dx.doi.org/10.1103/PhysRevLett.122.071801}{{\em Phys. Rev. Lett.} {\bf 122} (2019) no.~7, 071801}, \href{http://arxiv.org/abs/1806.03310}{{\tt arXiv:1806.03310 [hep-ph]}}.

\bibitem{ArgoNeuT:2019ckq}
{\bf ArgoNeuT} Collaboration, R.~Acciarri {\em et al.}, ``{Improved Limits on Millicharged Particles Using the ArgoNeuT Experiment at Fermilab},'' \href{http://dx.doi.org/10.1103/PhysRevLett.124.131801}{{\em Phys. Rev. Lett.} {\bf 124} (2020) no.~13, 131801}, \href{http://arxiv.org/abs/1911.07996}{{\tt arXiv:1911.07996 [hep-ex]}}.

\bibitem{Ball:2020dnx}
A.~Ball {\em et al.}, ``{Search for millicharged particles in proton-proton collisions at $\sqrt{s} = 13$ TeV},'' \href{http://dx.doi.org/10.1103/PhysRevD.102.032002}{{\em Phys. Rev. D} {\bf 102} (2020) no.~3, 032002}, \href{http://arxiv.org/abs/2005.06518}{{\tt arXiv:2005.06518 [hep-ex]}}.

\bibitem{Alcott:2025rxn}
S.~Alcott {\em et al.}, ``{Search for millicharged particles in proton-proton collisions at $\sqrt{s} = 13.6$ TeV},'' \href{http://arxiv.org/abs/2506.02251}{{\tt arXiv:2506.02251 [hep-ex]}}.

\bibitem{SENSEI:2023gie}
{\bf SENSEI} Collaboration, L.~Barak {\em et al.}, ``{Search by the SENSEI Experiment for Millicharged Particles Produced in the NuMI Beam},'' \href{http://dx.doi.org/10.1103/PhysRevLett.133.071801}{{\em Phys. Rev. Lett.} {\bf 133} (2024) no.~7, 071801}, \href{http://arxiv.org/abs/2305.04964}{{\tt arXiv:2305.04964 [hep-ex]}}.

\bibitem{Ilten:2022lfq}
P.~Ilten {\em et al.}, ``{Experiments and Facilities for Accelerator-Based Dark Sector Searches},'' in {\em {Snowmass 2021}}.
\newblock 6, 2022.
\newblock \href{http://arxiv.org/abs/2206.04220}{{\tt arXiv:2206.04220 [hep-ex]}}.

\bibitem{Alekhin:2015byh}
S.~Alekhin {\em et al.}, ``{A facility to Search for Hidden Particles at the CERN SPS: the SHiP physics case},'' \href{http://dx.doi.org/10.1088/0034-4885/79/12/124201}{{\em Rept. Prog. Phys.} {\bf 79} (2016) no.~12, 124201}, \href{http://arxiv.org/abs/1504.04855}{{\tt arXiv:1504.04855 [hep-ph]}}.

\bibitem{Ahdida:2023okr}
C.~Ahdida {\em et al.}, ``{Post-LS3 Experimental Options in ECN3},'' \href{http://arxiv.org/abs/2310.17726}{{\tt arXiv:2310.17726 [hep-ex]}}.

\end{thebibliography}\endgroup

\end{document}